\documentclass[journal]{IEEEtran}

\usepackage{tikz}
\usetikzlibrary{matrix}
\usetikzlibrary{calc}
\usepackage{booktabs} % For better table lines
\usepackage{listings}
\usepackage{xcolor}  % To customize colors

\usepackage{listings}
\definecolor{mybackground}{rgb}{1,1,1}  % White background
\definecolor{mytext}{rgb}{0,0,0}        % Black text
\tikzstyle{tensor}=[circle,draw=blue!50,fill=blue!20,thick]
\tikzstyle{tensor2}=[circle,draw=red!50,fill=red!20,thick]

\newcommand{\keywords}[1]{\vspace{0.5em}\noindent\textbf{Keywords:} #1}
\usepackage{amsmath}
\usepackage{amssymb}
\usepackage{siunitx}
\usepackage{mathtools}
\usepackage{hyperref}
\usepackage{algorithm}% http://ctan.org/pkg/algorithm
\usepackage{algpseudocode}% http://ctan.org/pkg/algorithmicx

\begin{document}
\title{A Matrix Product State Model for Simultaneous Classification and Generation}

\author{
  \IEEEauthorblockN{Alex Mossi\IEEEauthorrefmark{1}, Bojan Žunkovič\IEEEauthorrefmark{2}, Kyriakos Flouris\IEEEauthorrefmark{3}}
  \\
  \IEEEauthorblockA{\IEEEauthorrefmark{1}\textit{D-MATH}, \textit{ETH Zürich}, alex.mossi@alumni.ethz.ch} \\
  \IEEEauthorblockA{\IEEEauthorrefmark{2}\textit{FRI}, \textit{University of Ljubljana}, bojan.zunkovic@uni-lj.si} \\
  \IEEEauthorblockA{\IEEEauthorrefmark{3}\textit{D-ITET}, \textit{ETH Zürich}, kflouris@ethz.ch}
}

% \markboth{Journal of \LaTeX\ Class Files,~Vol.~14, No.~8, August~2015}%
% {Shell \MakeLowercase{\textit{et al.}}: Bare Demo of IEEEtran.cls for IEEE Journals}



\maketitle

\begin{abstract}

Quantum machine learning (QML) is a rapidly expanding field that merges the principles of quantum computing with the techniques of machine learning. One of the powerful mathematical frameworks in this domain is tensor networks. These networks are used to approximate high-order tensors by contracting tensors with lower ranks. Initially developed for simulating quantum systems, tensor networks have become integral to quantum computing and, by extension, to QML. 
Drawing inspiration from these quantum methods, specifically the Matrix Product States (MPS), we apply them in a classical machine learning setting. 
Their ability to efficiently represent and manipulate complex, high-dimensional data makes them effective in a supervised learning framework.
Here, we present an MPS model, in which the MPS functions as both a classifier and a generator. The dual functionality of this novel MPS model permits a strategy that enhances the traditional training of supervised MPS models. This framework is inspired by generative adversarial networks and is geared towards generating more realistic samples by reducing outliers. In addition, our contributions offer insights into the mechanics of tensor network methods for generation tasks.  Specifically, we discuss alternative embedding functions  and a new sampling method from non-normalized MPSs.
%First, to facilitate the compatibility of input data with MPS theory, an embedding function is introduced.
% We discuss the requisite properties of the MPS embedding function, providing a comparative analysis of prominent alternatives, including Fourier and Legendre embeddings. Performance assessments are conducted for generative tasks. The unique attributes of the aforementioned embedding functions allows for simultaneous use of MPS models for classification and data generation. . . Moreover, this study addresses issues associated with numerical instability that may arise during tensor contractions.
%It also delves into the latent space representation of generator-style MPS models and evaluates the impact of perturbations on classification accuracy across diverse embedding functions.
\end{abstract}

% % Note that keywords are not normally used for peerreview papers.
% \begin{IEEEkeywords}
% Tensor Networks, MPS, GAN, Machine Learning, 
% \end{IEEEkeywords}

\IEEEpeerreviewmaketitle

\section{Introduction}

In recent years, quantum technologies have undergone rapid and substantial advancements, holding the promise of revolutionizing various scientific \cite{PhysRevB.105.235122} and industrial domains~\cite{Ramakrishnan2023}. These advancements are closely intertwined with quantum machine learning (QML)~\cite{Suzuki2024, Hdaib2024}, and tensor networks (TNs) emerging as a vital mathematical tool. Central to the concept of TNs is their ability to approximate high-dimensional tensors via memory-efficient multi-dimensional arrays~\cite{tn:nutshell}.
% TNs provide a framework for representing and manipulating high-dimensional data or complex systems in an organized and computationally efficient manner~\cite{tn:nutshell}.Central to the concept of TNs is their utilization of multidimensional arrays to capture intricate relationships and to approximate higher-dimensional tensors.
%Central to the concept of TNs is their ability to approximate higher dimensional tensors via memory-efficient multidimensional arrays.
Originally devised as tools for the approximation and simulation of quantum systems within the domains of many-body quantum physics~\cite{mps:approximations} and condensed matter physics~\cite{tn:intro}, TNs now encompass a diverse array of fields, including data compression~\cite{mps:data_compression}, privacy~\cite{privacy}, and high-dimensional PDEs~\cite{pdes}. Of particular interest is their application in machine learning for both supervised~\cite{mps:supervised} and unsupervised tasks~\cite{mps:unsupervised,peps:supervised,peps:generative}.

% Diverse models of tensor networks have been conceived over time, each distinguished by the shape and properties of their constituent tensors, along with the specific rules and procedures followed when merging these tensors together through the process of contraction.
Prominent tensor networks in contemporary research include Projected Entangled Pair States (PEPS)~\cite{peps:generative}, Matrix Product State (MPS)~\cite{mps:supervised}~\cite{mps:unsupervised}, Locally Purified State (LPS)~\cite{lps}, Multiscale Entanglement Renormalization Ansatz (MERA)~\cite{mera}, tree tensor networks (TTN)~\cite{ttn}, and isometric tensor networks~\cite{isometric}. Of note, MPS, characterized by rank-3 tensors and sequential one-dimensional contractions, has garnered considerable attention due to its relative simplicity and versatility\cite{mpspeps:intro}.

MPSs were originally used to describe and simulate the quantum states of one-dimensional systems since they can faithfully represent quantum states featuring limited entanglement~\cite{mps:quantum1}~\cite{mps:quantum2}~\cite{mps:quantum3}. Such systems are notoriously challenging owing to the curse of dimensionality, which arises from the exponential growth of Hilbert spaces in such contexts~\cite{mps:curseofdim}. Since then, MPSs have been adapted to address a wide spectrum of datasets, including two-dimensional systems~\cite{mps:2dsystems}, rendering them suitable for image processing in machine learning applications. This has led to their successful application in tasks such as image classification using datasets like MNIST~\cite{mnist} and Fashion MNIST~\cite{fmnist}. While early training methods for MPSs in machine learning relied on density matrix renormalization group (DMRG) techniques~\cite{mps:supervised}, contemporary machine learning libraries provide more accessible approaches~\cite{mps:easier_training} based on automatic differentiation~\cite{diffprograming, stable}.

MPS models are typically employed for conventional classification~\cite{mps:supervised} or as generators in unsupervised contexts~\cite{mps:unsupervised}. However, the distinctive structure of MPS, combined with the characteristics of the embedding functions employed in this work, allows the use of a single model for both classification and generation tasks~\cite{bishop}\cite{NEURIPS2023_572a6f16}\cite{flouris2025}. This enables the use of a GAN-style method to improve its generative performance without affecting its classification accuracy.

We propose a novel approach for training supervised MPSs in a GAN-style setting. The MPS serves as both a classifier and a generator, resulting in improved generative performance and a reduction in the number of outliers generated, while maintaining robust classification accuracy. To this end, we present several contributions that allow the realization of such a model while providing insights into the mechanics of tensor network methods for generation. Thus, we are broadening the utility of tensor network methods for machine learning tasks.

The first part of this work establishes the theoretical framework underlying MPS models, as described in Sections.~\ref{sec:background} and ~\ref{sec:ml}. First, Matrix Product States are introduced, with a discussion of their canonical forms and why these forms are not strictly required in machine learning applications. Second, this section provides an overview of embedding functions, which are essential tools for transforming input data into representations compatible with MPS structures. Third, alternative embedding functions are explored, and techniques that simplify the computation of marginalized probability density functions (PDFs) over single variables, \( p(x_i) \), while avoiding the evaluation of high-dimensional integrals, are introduced. Fourth, an exact sampling procedure for non-normalized MPS is discussed, which removes the need for iterative sampling methods like Markov chain Monte Carlo (MCMC).

The second part of this work focuses on practical applications of MPS models in machine learning, as outlined in Sect.~\ref{sec:tensorgan}. The dual capability of MPS for simultaneous classification and generation tasks is emphasized, with particular attention to the GAN-style framework that serves as the foundation of this study. Key implementation challenges, such as managing exploding and vanishing values during tensor contractions, are addressed through specific techniques designed to stabilize training. Additionally, the use of MPS as a generator is examined, highlighting its ability to provide a latent space representation of the input data, which enables meaningful insights into its structure. The impact of perturbations within the embedded space on classification performance is also analyzed, with comparisons made across different embedding functions.

Finally, Sect.~\ref{sec:result} presents a detailed evaluation of our approach, including experimental results that validate the performance of GAN-style training and analyze the properties of the embedding functions. Sect.~\ref{sec:conclusion} then summarizes the findings and discusses potential directions for future research.

\section{Methods \label{sec:methods}
}
\subsection{Matrix product state and embedding functions}
\label{sec:background}
A Matrix Product State can be formally defined as a collection of $n-2$ rank-3 tensors $\{A_i^{\alpha_{i-1}, \alpha_{i}, d_i}\}_{i \in \{2, \ldots n-1\}}$ and two rank-2 tensors $\{A_1^{\alpha_1, d_1}, A_n^{\alpha_{n-1}, d_n}\}$, called sites, that can be contracted sequentially, resulting in the decomposition of a rank-$n$ tensor $W$, as described in the following equation:

\begin{multline}\label{eq:mps_def}
W^{d_1,d_2, ..., d_n}=\\
\sum_{\{\alpha_i\}} A_{1}^{\alpha_1, d_1} A_{2}^{\alpha_1 \alpha_2, d_2} \cdots A_{j}^{\alpha_{j-1} \alpha_j, d_j} \cdots A_{N}^{\alpha_{N-1}, d_N} .
\end{multline}
 The physical dimension $d$ corresponds to the dimension of indices $d_i$, and its role will be discussed in Sect.~\ref{subsec:embedding}. 
The bond indices \( \alpha_i \) determine the expressivity of the MPS, with the bond dimension \( D \) representing their maximum allowed size. Larger $D$ enables modeling of more complex correlations but increases computational cost, see \hyperref[app]{Appendix}. 

\begin{figure}[h!]
    \centering
    \begin{tikzpicture}[inner sep=0.5mm]
    \node[rectangle,draw=blue!50,fill=blue!20,thick, minimum width = 2cm, minimum height = .6cm] (0) at (-1, 0) {$W$};
    \draw[-] (-.2,-.3) -- (-.2,-.7);
    \draw[-] (-.6,-.3) -- (-.6,-.7);
    \node[-] at (-1,-.5) {$\ldots$};

    \draw[-] (-1.4,-.3) -- (-1.4,-.7);
    \draw[-] (-1.8,-.3) -- (-1.8,-.7);

    \node at (1,0) {=};

    \node[tensor, minimum size=2.3em] (1) at (2, 0) {\tiny$A_1$};
    \node (1a) at (2, -.7) {};
    \draw[-] (1) -- (1a);

    \node[tensor, minimum size=2.3em] (2) at (3, 0) {\tiny$A_2$};
    \node (2a) at (3, -.7) {};
    \draw[-] (2) -- (2a);

    \node (3) at (4, 0) {$\ldots$};

    \node[tensor, minimum size=2.0em] (4) at (5, 0) {\tiny$A_{n-1}$};
    \node (4a) at (5, -.7) {};
    \draw[-] (4) -- (4a);

    \node[tensor, minimum size=2.3em] (5) at (6, 0) {\tiny$A_n$};
    \node (5a) at (6, -.7) {};
    \draw[-] (5) -- (5a);

    \foreach \i in {1,...,2} {
        \pgfmathtruncatemacro{\iplusone}{\i + 1};
        \draw[-] (\i) -- (\iplusone);
    };

    \foreach \i in {3,...,4} {
        \pgfmathtruncatemacro{\iplusone}{\i + 1};
        \draw[-] (\i) -- (\iplusone);
    };
  
    \end{tikzpicture}
    \caption{Penrose diagram of the MPS decomposition in Eq. \ref{eq:mps_def}. Horizontal lines, bond indices \(\alpha_i\), connect adjacent tensors \(A_i\), while vertical lines, physical indices \(d_j\), represent input features via \(\phi (x_i)\). Boundary tensors \(A_1,A_n\) are rank-2, two open indices, and internal tensors are rank-3, three open indices}
    \label{fig:mps}
\end{figure}

A visual representation of an MPS decomposition in Eq.~\ref{eq:mps_def} is presented in Fig.~\ref{fig:mps}, using Penrose graphical notation for tensors~\cite{penrose}.
Each tensor $A_i$, represented by the blue circles in Fig.~\ref{fig:mps}, has horizontal indices $\alpha_i$, whose size corresponds to the bond dimension $D$, and vertical indices $d_i$, whose size corresponds to the physical dimension $d$. The boundary tensors $A_1$ and $A_n$ are rank-2, while intermediate tensors $A_2,\ldots,A_n$ are rank-3.
%The MPS decomposition in Eq.~\ref{eq:mps_def} corresponds directly to the graphical representation in Fig.~\ref{fig:mps}.

Throughout this work, we use different Penrose tensor network representations of the MPS depending on the context. In some diagrams, such as Fig.~\ref{fig:mps}, we explicitly show the tensors \( A_i \) as separate entities, highlighting the individual components of the MPS. However, in later TN diagrams, such as in Eq.~\ref{eq:mpspdf}, we omit the explicit labeling of the sites \( A_i \) for clarity, as the focus shifts toward the overall network structure rather than the individual tensors. For completeness, a detailed description of all equations involving Penrose tensor notation is provided in \hyperref[app]{Appendix}, where each expression is explicitly rewritten in standard indexed notation.

Additionally, the TN representation changes when the MPS is contracted with an input. In the initial formulation, vertical edges represent the physical indices \( d_j \), which correspond to the input space. These are referred to as open indices, as they are not initially contracted with other tensors. When the MPS is contracted with the embedded input \( \Phi(\mathbf{x}) \), these vertical edges become connected to the one-dimensional tensor \( \Phi(\mathbf{x}) \), effectively integrating the input into the network.

% Throughout this work, we use different Penrose tensor network representations of the MPS depending on the context. In some diagrams, such as Fig.~\ref{fig:mps}, we explicitly show the tensors  $A_i$  as separate entities, highlighting the individual components of the MPS. However, in later TN diagrams, as for example Eq.~\ref{eq:mpspdf}, we omit the explicit labeling of the sites  $A_i$  for clarity, as the focus shifts toward the overall network structure rather than the individual tensors. For completeness, a detailed description of all equations involving Penrose tensor notation is provided in App.~\ref{app:penrose}, where each expression is explicitly rewritten in standard indexed notation.

% Additionally, the TN representation changes when the MPS is contracted with an input. In the initial formulation, vertical edges represent the physical indices  $d_j$ , which correspond to the input space. These are referred to as open indices, as they are not initially contracted with other tensors. When the MPS is contracted with the embedded input $\phi(x)$, these vertical edges become connected to the one-dimensional tensor $\phi(x)$, effectively integrating the input into the network.

An MPS can be used to approximate any rank-n tensor W using a DMRG-based method~\cite{DMRG}\cite{DMRG2}\cite{DMRG3} to find its optimal decomposition. In quantum physics, MPS approximates the wavefunction $\psi(\mathbf{x})$ of a system, allowing the calculation of the probability density as the squared magnitude: $P(\mathbf{x}) = |\psi(\mathbf{x})|^2$.
 Similarly, in our machine learning framework, the MPS is used to approximate the PDF of the data:

\begin{multline} \label{eq:mpspdf}
p(x_1, \ldots, x_n) = \\
\left[ \begin{tikzpicture}[inner sep=0.5mm, baseline=(current bounding box.center)]
    \foreach \i/\label in {1/x_1,2/x_2,3/\ldots,4/x_{n-1},5/x_n} {
        \ifnum\i=3
        \node[font=\footnotesize] (x\i) at (\i, 0) {$\ldots$};
        \node[font=\footnotesize] (y\i) at (\i, -.5) {};
        \node[font=\footnotesize] (y\i*) at (\i, -1) {};
        \else
        \node[tensor, font=\footnotesize] (x\i) at (\i, 0) {};
        \node[tensor, font=\footnotesize] (y\i) at (\i, -.5) {};
        \draw[-] (x\i) -- (y\i);
        \node[font = \footnotesize] at (\i, -.75) {$\phi(\label)$};
\fi
    }
    \foreach  \i in {1,...,4} {
        \pgfmathtruncatemacro{\next}{\i + 1}
        \pgfmathtruncatemacro{\nextstar}{\i + 1}
        \draw[-] (x\i) -- (x\next);
    }
\end{tikzpicture} \right] ^2 = \\
   \begin{tikzpicture}[inner sep=0.5mm, baseline=(current bounding box.center)]
    \foreach \i/\label in {1/x_1,2/x_2,3/\ldots,4/x_{n-1},5/x_n} {
        \ifnum\i=3
        \node[font=\footnotesize] (x\i) at (\i, 0) {$\ldots$};
        \node[font=\footnotesize] (x\i*) at (\i, -2) {$\ldots$};
        \node[font=\footnotesize] (y\i) at (\i, -.5) {};
        \node[font=\footnotesize] (y\i*) at (\i, -1) {};
        \node[font = \footnotesize] at (\i, -1) {$\label$};
        \else
        \node[tensor, font=\footnotesize] (x\i) at (\i, 0) {};
        \node[tensor, font=\footnotesize] (x\i*) at (\i, -2) {};
        \node[tensor, font=\footnotesize] (y\i) at (\i, -.5) {};
        \node[tensor, font=\footnotesize] (y\i*) at (\i, -1.5) {};
        \draw[-] (x\i) -- (y\i);
        \draw[-] (x\i*) -- (y\i*);
        \node[font = \footnotesize] at (\i, -.75) {$\phi(\label)$};
        \node[font = \footnotesize] at (\i, -1.25) {$\phi(\label)$};
\fi
    }
    \foreach  \i in {1,...,4} {
        \pgfmathtruncatemacro{\next}{\i + 1}
        \pgfmathtruncatemacro{\nextstar}{\i + 1}
        \draw[-] (x\i) -- (x\next);
        \draw[-] (x\i*) -- (x\nextstar*);
    }
\end{tikzpicture}.
\end{multline}

The use of the squared value \( |W(\mathbf{x})|^2 \) aligns with the generative interpretation of MPS, inspired by its origins in quantum mechanics, where probabilities are derived from the squared magnitudes of wavefunctions~\cite{mps:supervised}. This formulation ensures that the MPS framework remains both expressive and interpretable for probabilistic modeling.

We employ an embedding function, \( \Phi(\mathbf{x}) \), also referred to as a feature map or feature embedding in the tensor network literature~\cite{vzunkovivc2024grokking}. The embedding function transforms input vectors of length \( N \), with support \( \text{supp}(\Phi) = [0,1]^{N} \), into a format suitable for MPS theory.
 Since an MPS operates linearly on its input, through a sequence of tensor contractions and matrix multiplications, the embedding function introduces the necessary non-linearity into the system. 

The embedding function is defined as a tensor contraction of local feature maps \( \phi(x_i) \in \mathbb{R}^d \), which can alternatively be referred to as local embedding functions or local embedding features. Specifically:
\begin{equation}
\Phi(\mathbf{x}) = \phi(x_1) \otimes \dots \otimes \phi(x_n),
\end{equation}
and the specific forms of the feature maps are elaborated upon in Sect.~\ref{subsec:embedding}, where the necessary conditions are explained.

There are three crucial hyperparameters used to define the structure of an MPS. The first one is the number of sites, corresponding to the number of bodies in the quantum physics scenario, and is determined by the input size in our model settings and will be denoted by $N$. The second is the bond dimension, $D$, which governs the size of indices connecting two different sites. The third parameter is the physical dimension, $d$, and is equal to the dimension of the vector $\phi(x_i)$, itself determined by our choice of embedding functions. Our implementation of an MPS is stored in a single tensor with dimensions $(N, D, D, d)$. It should be noted that the boundary sites are also included in the tensor, although they possess fewer indices compared to the intermediate sites. This setup is called open boundary condition, used for non-periodic quantum systems~\cite{mps:quantum1}. For periodic quantum systems, where translational invariance is present, a different approach is employed where the first and last sites are rank-3 tensors connected to each other. However, this approach is usually not employed in the case of MPS applications in machine learning scenarios~\cite{mps:supervised, mps:unsupervised}.

\subsubsection{Canonical form of an MPS}
\label{subsec:canonicalization}
It is evident that an MPS does not possess a unique form. In fact, if we introduce any sequence of invertible matrices $\{X_i\}_{i \in \{1,...N-1\}}$ of size $d\times d$ between any two sites along with their inverses, and setting $X_0 = X_N = I_d$, we can transform the tensors that comprise the MPS as follows:

\begin{equation}
B_i = X_{i-1}^{-1} A_i X_i .
\end{equation}
This implies, ignoring the nonrelevant indices,

\begin{align}
W &= \sum_{\{\alpha\}} A_{1}^{\alpha_1} A_{2}^{\alpha_1 \alpha_2}  \cdots A_{N}^{\alpha_{N-1}} = \\
&= \sum_{\{\alpha\}} A_{1}^{\alpha_1} X_1 X_1^{-1} A_{2}^{\alpha_1 \alpha_2} X_2  \cdots X_{n-1}^{-1} A_{N}^{\alpha_{N-1}} = \\
&= \sum_{\{\alpha\}} B_{1}^{\alpha_1} B_{2}^{\alpha_1 \alpha_2} \cdots B_{j}^{\alpha_j \alpha_{j+1}} \cdots B_{N}^{\alpha_{N-1}},
\end{align}
proving the non-uniqueness of the representation of an MPS. These degrees of freedom are termed gauge degrees of freedom. The non-unique representation of MPSs is exploitable and can help devise better privacy-preserving machine learning algorithms~\cite{mps:privacy_preserving}. Canonical forms are also crucial for an efficient time-dependent variational principle for MPS ~\cite{haegeman2016unifying}.

To establish a canonical form for the MPS, various methods can be employed. One method uses DMRG-based and singular value decomposition techniques~\cite{DMRG}. Another alternative~\cite{canonicalform} employs the QR decomposition of the matrices that compose the MPS to achieve an orthogonal decomposition of the matrices. Those canonical forms of MPSs are needed to describe a so-called sweeping algorithm used for optimization~\cite{sweeping}. However, canonical forms can often be difficult to obtain for arbitrary MPSs~\cite{mps:tt_decomp}, since some of the calculations needed (for example, calculating the optimal rank $d$) are NP-hard~\cite{tn:np-complete} and/or an ill-posed problem~\cite{tn:ill-posed}.
% , but, thanks to the capabilities of torch automatic differentiation~\cite{torch:autograd}, a more straightforward approach can be implemented.
The sweeping algorithm and its necessary calculation of a canonical form will not be required as we can effectively leverage a more accessible approach via PyTorch automatic differentiation~\cite{torch:autograd}, discussed in Sect.~\ref{subsec:classification}.  Specifically, we optimize the parameters of the model by minimizing the cross-entropy loss function, an approach commonly used for tensor networks applied in machine learning scenarios~\cite{mps:easier_training}.

\subsubsection{Embedding function}
\label{subsec:embedding}
If the objective is data classification, the embedding function should primarily introduce non-linearity and transform the data such that our MPS 
%(which in itself, without the presence of the embedding function, is just a linear model) 
can effectively achieve linear separation in the high-dimensional space where the data is embedded. Conversely, if the model is being used for data sampling, there is a greater emphasis on the selection of the embedding function. To employ the methodology outlined in Sect.~\ref{subsec:generation} and perform simultaneous classification and generation tasks, an essential prerequisite for the  local embedding function $\phi(x_i)\in \mathbb{R}^d$ is as follows:
\begin{equation}
\label{eq:emb}
\int_{x_i\in X} \phi_j(x_i) \phi_k(x_i) dx_i = \delta_{j, k},
\end{equation}

where $X=[0,1]$, assumed to be the support of the input data, with limited exceptions noted herein. Eq.~\ref{eq:emb} allows for a greatly simplified computation of the marginal probability over single variables, avoiding high-dimensional integrals as shown in Sect.~\ref{subsec:density_matrix}.
% becomes necessary due to the fact that, when computing marginal probabilities via integration over the variables $x_i$, we shall leverage this characteristic in our computations, and in Sect.~\ref{subsec:density_matrix} we described how we exploit this fact during the computation of the marginal probability over single variables to avoid the computation of high dimensional integrals. It is noteworthy that, 
In certain cases, the embedding functions may also be defined with

\begin{equation}
\int_{x_i\in X} \phi_j(x_i)\phi_k(x_i)dx_i = c \cdot \delta_{j,k}, \quad c>0 .
\end{equation}

Such a condition yields a non-normalized PDF. Nevertheless, we demonstrate that our method can accommodate any resulting inconsistencies, Sect.~\ref{subsec:generation},~\ref{subsec:normalization}.

% However, this variation does not affect our methodology. Instead, it yields a non-normalized PDF, but, as discussed in Sect.~\ref{subsec:generation} and~\ref{subsec:normalization}, we will deal with non-normalized probabilities density functions all throughout this work.

% As delineated in prior investigations~\cite{mps:supervised}, when the functions $\{\phi_s(x)\}, s = 1, 2, \ldots, d$ constitute a basis for a Hilbert space of functions defined over the domain $x \in [0, 1]$, it follows that the tensor product basis

% \begin{equation}
% \phi_1(x_1) \otimes \phi_2(x_2) \otimes \ldots \otimes \phi_n(x_n)
% \end{equation}
% forms a basis for a Hilbert space of functions defined over the domain $x \in [0, 1]^n$. Furthermore, if the basis $\{\phi_s(x)\}$ is deemed complete, then the tensor product basis also attains completeness, thereby encompassing the capacity to represent any square-integrable function, denoted as $f(x)$.

The marginal probability over a single variable will have the form of
\begin{equation}
\label{eq:pdf}
    P(x_i) = \phi(x_i)\cdot V_i \cdot \phi(x_i) ,
\end{equation}
with $V_{i}$ being the symmetric and semi-positive definite reduced density matrix, Sect.~\ref{subsec:density_matrix}. This implies that the PDF and its complexity will depend on our choice of $\phi(x_i)$ and the physical dimension of the model. 

In the case of handling simple low-frequency data, such as high-contrast or binary images, a common embedding function~\cite{peps:supervised}~\cite{mps:supervised} is given by the following:

\begin{equation}
\phi(x_i) = [\sin(\frac{\pi}{2} x_i), \cos(\frac{\pi}{2} x_i)] .
\end{equation}

This adheres to the constraint specified in Eq.~\ref{eq:emb} over the interval $[-1, 1]$. However, this function exhibits a limitation in that it possesses a low physical dimension of only 2. Consequently, for certain datasets, it may struggle to introduce the requisite complexity, thereby potentially impeding the model's performance.

A generalized replacement, outlined in~\cite{mps:supervised}, involves the use of the embedding function:

\begin{equation} \label{eq:genemb}
    \phi_j(x_i) = \sqrt{\binom{d-1}{j-1}} \cos(x_i)^{d-j} \sin(x_i)^j .
\end{equation}

This function belongs to the class of functions referred to as spin coherent states~\cite{mps:supervised}. However, Eq.~\ref{eq:genemb} does not satisfy the condition specified in Eq.~\ref{eq:emb} and is therefore not a candidate embedding function for generation.
% Consequently, in our study, where the aim is to leverage the generative capabilities of MPSs, we refrain from using this embedding function. Its non-conformance with Eq.~\ref{eq:emb} would preclude us from employing the method described in Sect.~\ref{subsec:generation}.

%, even though for larger d the components of $\phi_i(x_j)$, it contains ever higher frequency terms of the form $cos(\pi/2 x_j), cos(3\pi/2 x_j), ...,$ this function requires a higher d than our choice in order to obtain the same complexity, and a better alternative is to use directly the so-called Fourier embedding:

An alternative proposal is the Fourier embedding ~\cite{mps:positive_unlabeled_learning}:

\begin{equation}
\label{eq:fourier}
    \phi_j(x_i) = \text{cos}(j \pi x_i) \text{, for } j \in \{0,...d-1\} ,
\end{equation}

for $\text{Supp}(\Phi)=[0,1]^n$.
This feature map satisfies Eq.~\ref{eq:emb} and can model more general PDFs, due to the fact that we can use arbitrarily high values of $d$. In this case, the PDF over a single variable will be modeled by the following equation:

\begin{equation}
\sum_{j = 0}^{d-1}\sum_{k = 0}^{d-1} a_{jk} \text{cos}(\frac{\pi}{2}jx_i) \text{cos}(\frac{\pi}{2}kx_i).
\end{equation}

% We'll explore the use of other embedding function, not used until now in the MPS framework. 

Our alternative proposal for an embedding function for high-dimensional physical spaces is to leverage polynomials, and due to Eq.~\ref{eq:emb}, a natural choice is the use of Legendre polynomials. These are a set of polynomials 
$\{P_1(x_i), P_2(x_i), ...\}$ such that $P_j(x_i)$ is a polynomial of degree $j$ and any 2 polynomials satisfy Eq.~\ref{eq:emb}, obtained recursively by
\begin{align}
    P_0(x_i) &= 1, \\
    P_1(x_i) &= x_i, \\
    P_j(x_i) &= \frac{(2j - 1)x_iP_{j-1} - (j - 1)P_{n-2}(x_i)}{j}, \quad \forall j \geq 2 .
\end{align}

Unliked Fourier embeddings, $\text{Supp}(P)=[-1,1]^N$ instead of the previous interval $[0,1]^N$, so $P$ will require rescaling in a pre-processing step. This embedding leads to a PDF over a single variable which is modelled by a non-negative polynomial of degree $d^2$ over $[-1,1]$.

Fourier and Legendre embeddings can be used with arbitrarily large physical dimensions, thus suitable for modelling multi-modal probability distribution functions during the generative phase of Sect.~\ref{subsec:generation}. The generative results of these embedding functions are discussed in Sect.~\ref{sec:result}.

\subsubsection{Computing the reduced density matrix}
\label{subsec:density_matrix}
% The calculations for the reduced density matrix in~\cite{unitary_tn:sampling} are described in the case of unitary tensor networks, we will have to use a slightly different approach. 

Given any local embedding function $\phi(x_i)$, we define the following matrix:
\begin{equation}\label{eq:bdef}
    B \coloneqq \int \begin{tikzpicture}[inner sep=0.5mm, baseline={(current bounding box.center)}]
        \node[tensor] (x) at (0, -0.5) {};
        \node[tensor] (x*) at (0, 0.5) {};
        \node[font = \footnotesize] at (0, 0.25) {$\phi(x_i)$};
        \node[font = \footnotesize] at (0, -0.25) {$\phi(x_i)$};
        \draw[-] (0,-0.75) -- (x);
        \draw[-] (0,0.75) -- (x*);
    \end{tikzpicture} dx_i.
\end{equation}
This is in contrast to~\cite{unitary_tn:sampling}, where the reduced density matrix is described by unitary tensor networks. Recalling that the PDF is approximated by an MPS by Eq.~\ref{eq:mpspdf}, the marginalized PDF can be simplified over a single variable $p(x_i)$ without necessitating the evaluation of multi-dimensional integrals:
% e can now observe that, starting from the joint probability distribution $p(x_1, \ldots, x_n)$, we can simplify the marginalized PDF over a single variable $p(x_i)$, without necessitating the evaluation of multi-dimensional integrals:
\begin{multline}\label{eq:cond_prob1}
p(x_i) = 
\\
\int \ldots \int
\begin{tikzpicture}[inner sep=0.5mm, baseline=(current bounding box.center)]
    \foreach \i/\label in {1/x_1,2/\ldots,3/x_n} {
        \ifnum\i=2
            \node[font=\footnotesize] (x\i) at (\i, 1) {$\ldots$};
            \node[font=\footnotesize] (x\i*) at (\i, -1) {$\ldots$};
            \node[font=\footnotesize] (y\i) at (\i, .5) {};
            \node[font=\footnotesize] (y\i*) at (\i, -0.5) {};
            \node[font = \footnotesize] at (\i, 0) {$\label$};
        \else    
                \node[tensor, font=\footnotesize] (x\i) at (\i, 1) {};
                \node[tensor, font=\footnotesize] (x\i*) at (\i, -1) {};
                \node[tensor, font=\footnotesize] (y\i) at (\i, 0.5) {};
                \node[tensor, font=\footnotesize] (y\i*) at (\i, -0.5) {};
                \draw[-] (x\i) -- (y\i);
                \draw[-] (x\i*) -- (y\i*);
                % \node[font = \footnotesize] at (\i, -.75) {$\phi(\label)$};
                \node[font = \footnotesize] at (\i, 0.25) {$\phi(\label)$};
                \node[font = \footnotesize] at (\i, -0.25) {$\phi(\label)$};
        \fi
    }
    \foreach  \i in {1,...,2} {
        \pgfmathtruncatemacro{\next}{\i + 1}
        \pgfmathtruncatemacro{\nextstar}{\i + 1}
        \draw[-] (x\i) -- (x\next);
        \draw[-] (x\i*) -- (x\nextstar*);
    }
\end{tikzpicture}
dx_1,\ldots \widehat{dx_i}, \ldots ,dx_n =
\\
\begin{tikzpicture}[inner sep=0.5mm, baseline=(current bounding box.center)]
    % \foreach \i/\label in {1/x_1,2/\ldots,3/x_{i-1},4/x_i,5/x_{i+1}, 6/\ldots, 7/x_n} {
    \foreach \i/\label in {1/x_1,2/\ldots,3/x_i, 4/\ldots, 5/x_n} {
        \ifnum\i=3
            \node[tensor, font=\footnotesize] (x\i) at (\i, 1) {};
            \node[tensor, font=\footnotesize] (x\i*) at (\i, -1) {};
            \node[tensor, font=\footnotesize] (y\i) at (\i, 0.5) {};
            \node[tensor, font=\footnotesize] (y\i*) at (\i, -0.5) {};
            \draw[-] (x\i) -- (y\i);
            \draw[-] (x\i*) -- (y\i*);
            \node[font=\footnotesize] at (\i, 0.25) {$\phi(x_i)$};  
            \node[font=\footnotesize] at (\i, -0.25) {$\phi(x_i)$};       
        \else
            \ifnum\i=2
                \node[font=\footnotesize] (x\i) at (\i, 1) {$\ldots$};
                \node[font=\footnotesize] (x\i*) at (\i, -1) {$\ldots$};
                \node[font=\footnotesize] (y\i) at (\i, 0.5) {};
                \node[font=\footnotesize] (y\i*) at (\i, -0.5) {};
                \node[font=\footnotesize] at (\i, 0) {$\label$};
            \else
                \ifnum\i=4
                    \node[font=\footnotesize] (x\i) at (\i, 1) {$\ldots$};
                    \node[font=\footnotesize] (x\i*) at (\i, -1) {$\ldots$};
                    \node[font=\footnotesize] (y\i) at (\i, 0.5) {};
                    \node[font=\footnotesize] (y\i*) at (\i, -0.5) {};
                    \node[font=\footnotesize] at (\i, 0) {$\label$};
                \else
                    \node[tensor, font=\footnotesize] (x\i) at (\i, 1) {};
                    \node[tensor, font=\footnotesize] (B) at (\i, 0) {$B$};
                    \node[tensor, font=\footnotesize] (x\i*) at (\i, -1) {};
                    \draw[-] (x\i) -- (B); 
                    \draw[-] (B) -- (x\i*); 
                \fi
            \fi
        \fi
    }
    \foreach  \i in {1,...,4} {
        \pgfmathtruncatemacro{\next}{\i + 1}
            \draw[-] (x\i) -- (x\next);
            \draw[-] (x\i*) -- (x\next*);
    }
\end{tikzpicture} .
\end{multline}

We assume that the selection of \( \phi(x) \) satisfies Eq.~\ref{eq:emb} such that \( B = \mathbb{I}_d \), leading to

% \begin{equation}
% \int \begin{tikzpicture}[inner sep=0.5mm, baseline={(current bounding box.center)}]
%         \node[tensor] (x) at (0, -0.5) {};
%         \node[tensor] (x*) at (0, 0.5) {};
%         \node[font = \footnotesize] at (0, 0.25) {$\phi(x)$};
%         \node[font = \footnotesize] at (0, -0.25) {$\phi(x)$};
%         \draw[-] (0,-0.75) -- (x);
%         \draw[-] (0,0.75) -- (x*);
%     \end{tikzpicture} dx = \mathbb{I}_d 
%     \iff 
%     \int \phi_i(x)\phi_j(x) = \delta_{i,j} ,  \quad \forall i,j \in \{1,..,d\} ,
% \end{equation} 

\begin{multline}
\label{eq:marginalize}
p(x_i) =  \\
\begin{tikzpicture}[inner sep=0.5mm, baseline=(current bounding box.center)]
    % \foreach \i/\label in {1/x_1,2/\ldots,3/x_{i-1},4/x_i,5/x_{i+1}, 6/\ldots, 7/x_n} {
    \foreach \i/\label in {1/x_1,2/\ldots,3/x_i, 4/\ldots, 5/x_n} {
        \ifnum\i=3
            \node[tensor, font=\footnotesize] (x\i) at (\i, 1) {};
            \node[tensor, font=\footnotesize] (x\i*) at (\i, -1) {};
            \node[tensor, font=\footnotesize] (y\i) at (\i, 0.5) {};
            \node[tensor, font=\footnotesize] (y\i*) at (\i, -0.5) {};
            \draw[-] (x\i) -- (y\i);
            \draw[-] (x\i*) -- (y\i*);
            \node[font=\footnotesize] at (\i, 0.25) {$\phi(x_i)$};  
            \node[font=\footnotesize] at (\i, -0.25) {$\phi(x_i)$};       
        \else
            \ifnum\i=2
                \node[font=\footnotesize] (x\i) at (\i, 1) {$\ldots$};
                \node[font=\footnotesize] (x\i*) at (\i, -1) {$\ldots$};
                \node[font=\footnotesize] (y\i) at (\i, 0.5) {};
                \node[font=\footnotesize] (y\i*) at (\i, -0.5) {};
                \node[font=\footnotesize] at (\i, 0) {$\label$};
            \else
                \ifnum\i=4
                    \node[font=\footnotesize] (x\i) at (\i, 1) {$\ldots$};
                    \node[font=\footnotesize] (x\i*) at (\i, -1) {$\ldots$};
                    \node[font=\footnotesize] (y\i) at (\i, 0.5) {};
                    \node[font=\footnotesize] (y\i*) at (\i, -0.5) {};
                    \node[font=\footnotesize] at (\i, 0) {$\label$};
                \else
                    \node[tensor, font=\footnotesize] (x\i) at (\i, 1) {};
                    \node[tensor, font=\footnotesize] (x\i*) at (\i, -1) {};
                    \draw[-] (x\i) -- (x\i*); 
                \fi
            \fi
        \fi
    }
    \foreach  \i in {1,...,4} {
        \pgfmathtruncatemacro{\next}{\i + 1}
            \draw[-] (x\i) -- (x\next);
            \draw[-] (x\i*) -- (x\next*);
    }
\end{tikzpicture} .
\end{multline}

The marginalized distribution over a single variable is given by Eq.~\ref{eq:pdf}. 
% holds and that the marginalized distribution over a single variable can also be written as

% \begin{equation}
%     p(x_i) = \phi(x_i) \cdot V_i \cdot \phi(x_i)
% \end{equation}
% where $V_i$ is the previously cited reduced density matrix, which can be generalized also 
In the case where we want to calculate a conditional probability given some other variables' value, we can substitute the value of the conditioning variables. %instead of contracting the two copies of the MPS.
This conditional probability density over a single variable is given by the following:
\begin{equation}
    p(x_i|\{x_j\}_{j\in \mathcal{I}}) = \phi(x_i) \cdot V_{i, \{x_{j}\}_{j\in \mathcal{I}}} \cdot \phi(x_i).
\end{equation}
In order to compute the density matrix $V_{i, \{x_{j}\}_{j\in \mathcal{I}}}$, for $\{x_j| j \in \mathcal{I}\}$, we perform a contraction operation involving two copies of the MPS. At each site, there are three possible scenarios for the tensor contraction in the physical dimension:

\begin{enumerate}
\label{list:density_matrix}
    \item For $j = i$, we leave these two indices open.
    \item For $j \in \mathcal{I}$, we compute the embedding $\phi(x_j)$ and contract the physical indices at site $j$ of the two copies of the MPS with $\phi(x_j)$.
    \item   For $j \notin \mathcal{I}$, we directly contract the open indices of the two copies of the MPS, as described visually in Eq.~\ref{eq:marginalize}.
\end{enumerate}

The outcome of this function is a semi-positive definite symmetric matrix, a real matrix $M$ is semi-positive definite if and only if there exists a matrix $B$ such that $M=B^\intercal B$. Without loss of generality and supposing $\mathcal{I} = \{1, \cdots, i-1\}$, $B$ is defined as
%TODO: correct this
\begin{equation}\label{eq:b_def}
    B_i \coloneqq
\begin{tikzpicture}[inner sep=0.2mm, baseline={(current bounding box.center)}]
    \node[tensor, minimum size=2.0em] (x1) at (1, 0) {\tiny $A_1$};
    \node[tensor, font=\footnotesize] (y1) at (1, -.75) {};
    \draw[-] (x1) -- (y1);
    \node[font=\footnotesize] at (1, -1) {$\phi(x_1)$};

    \node[font=\footnotesize] (x2) at (2, 0) {$\ldots$};
    \node[font=\footnotesize] (y2) at (2, -.5) {};
    \node[font=\footnotesize] at (2, -.75) {};
    
    \node[tensor, font=\footnotesize, minimum size=2.0em] (x3) at (3, 0) {\tiny$A_{i-1}$};
    \node[tensor, font=\footnotesize] (y3) at (3, -.75) {};
    \draw[-] (x3) -- (y3);
    \node[font=\footnotesize] at (3, -1) {$\phi(x_{i-1})$};

    \node[tensor, font=\footnotesize, minimum size=2.0em] (x4) at (4, 0) {\tiny$A_i$};
    \node (y4) at (4, -0.75) {};
    \draw[-] (x4) -- (y4); 

    \node[tensor, font=\footnotesize, minimum size=2.0em] (x5) at (5, 0) {\tiny$A_{i+1}$};
    \node (y5) at (5, -0.75) {};
    \draw[-] (x5) -- (y5); 

    \node[font=\footnotesize] (x6) at (6, 0) {$\ldots$};
    \node[font=\footnotesize] (y6) at (6, -.5) {};
    \node[font=\footnotesize] at (6, -.75) {};

    \node[tensor, font=\footnotesize, minimum size=2.0em] (x7) at (7, 0) {\tiny$A_n$};
    \node (y7) at (7, -0.75) {};
    \draw[-] (x7) -- (y7); 
    % \foreach \i/\label in {1/x_1,2/\ldots,3/x_i,4/\ldots,5/x_n} {

            % \ifnum\i=3
            %     \node[font=\footnotesize] (x\i) at (\i, 0) {$\ldots$};
            %     \node[font=\footnotesize] (y\i) at (\i, -.5) {};
            %     \node[font=\footnotesize] at (\i, -.75) {$\label$};
            % \else
            %     \node[tensor, font=\footnotesize] (x\i) at (\i, 0) {};
            %     \node[tensor, font=\footnotesize] (x\i*) at (\i, -0.5) {};
            %     \draw[-] (x\i) -- (x\i*); 
            % \fi
    
    \foreach  \i in {1,...,6} {
        \pgfmathtruncatemacro{\next}{\i + 1}
            \draw[-] (x\i) -- (x\next);
    }
\end{tikzpicture}.
\end{equation}
After reformatting $B_i^{d_i, d_{i+1}, \ldots d_n}$ into matrix form $B_i'$ of shape $d^{n-i-1}\times d$, where we let the first index of the matrix correspond to the physical index of $A_i$ and we regroup the remaining physical indices of $\{A_{i+1}, ..., A_{n}\}$ as the second index of the matrix, the tensor contraction
$\sum_{j_{i+1}, ..., j_{n}} B_{j_1, ..., j_{i-1}}, B_{j_1, ..., j_{i-1}}$
will correspond to the matrix multiplication
$B_i'^\intercal B_i'$. We can rewrite $V_i = B_i'B_i'^\intercal$, proving its semi-positive definiteness.

Throughout this section, we did not assume to have a normalized MPS since that will not be the case during the training procedure. This can cause the computed PDFs to be unnormalized, but this will not affect our methods of classification and sampling, as we will see later in Sect.~\ref{subsec:generation}.

\subsection{Classification and generation}
\label{sec:ml}
\subsubsection{Classification with MPS}
\label{subsec:classification}
In this section, we will discuss two methods that are commonly employed to perform classification using MPSs~\cite{mps:supervised}. 
The first approach involves creating an ensemble of MPSs, where each MPS corresponds to a distinct label class within the data. This ensemble model generates an output vector of length $C$, where each component of the vector arises from the contraction of a different MPS with the input. In Fig.~\ref{fig:classification_ensemble} we show these contraction steps for a single component of the ensemble.

\begin{figure}[h!]
\centering
    \begin{tikzpicture}[inner sep=0.5mm]
        \foreach \i in {1,...,4} {
        
            \node[tensor, minimum size = 2.0em] (\i) at (\i, 0) {$A_\i^C$};
            \node[tensor] (x\i) at (\i, -1){};
            \node[font = \footnotesize] at (\i, -1.35) {$\phi(x_\i)$};
            \draw[red, -] (\i) -- (x\i);
        };
        % horizontal lines
        \foreach \i in {1, 2, 3} {
            \pgfmathtruncatemacro{\iplusone}{\i + 1};    
            \draw[-] (\i) -- (\iplusone);
        };
    \end{tikzpicture}
    \linebreak 
    \linebreak
    \begin{tikzpicture}[inner sep=0.5mm]
        \node[tensor, minimum size = 2.0em] (1) at (1, 0) {$L^C$};
        \node[tensor, minimum size = 2.0em] (2) at (2, 0) {$A_2^C$};
        \node[tensor, minimum size = 2.0em] (3) at (3, 0) {$A_3^C$};
        \node[tensor, minimum size = 2.0em] (4) at (4, 0) {$R^C$};
        % horizontal lines 
        \draw[red, -] (1) -- (2);
        \draw[-] (2) -- (3);
        \draw[red, -] (3) -- (4);
    \end{tikzpicture}
    \linebreak
    \linebreak 
    \begin{tikzpicture}[inner sep=0.5mm]
        \node[tensor, minimum size = 2.0em] (1) at (1, 0) {$L^C$};
        \node[tensor, minimum size = 2.0em] (2) at (2, 0) {$R^C$};        
        \draw[red, -] (1) -- (2);
    \end{tikzpicture}
    \linebreak
    \linebreak
    \begin{tikzpicture}[inner sep=0.5mm]
        % output tensor
        \node[tensor, minimum size = 2.0em] (c) at (1,0){$y^C$};
    \end{tikzpicture}
        
    \caption{This illustrates an example of how a single MPS of the ensemble, corresponding to the class $C$, and input are contracted in the forward pass, considering the case $N=4$. The red lines indicate the indices that are being contracted in each step. By squaring the value of the final scalar $y^C$, we get a non-normalized probability, i.e., $p(c=i|\mathbf{x}) = \frac{1}{Z}\cdot y_i^2 $, with $Z = \sum_c y_c^2$ being a normalization constant depending on the outputs of the ensemble of MPSs.}
    \label{fig:classification_ensemble}
\end{figure}

The second approach utilizes a single modified MPS, where we introduce an additional tensor placed in the middle of the tensor network. This supplementary tensor, with dimensions $(C, D, D)$, plays a crucial role in producing the desired vector of length $C$. The network structure of this approach is described by the following equation:

\begin{multline}
    W^{d_1,d_2, ..., d_n}_c=\\
    \sum_{\alpha} A_{1}^{\alpha_1 d_1} A_{2}^{\alpha_1 \alpha_2 d_2} \cdots C^{\alpha_{n/2} \alpha_{n/2+1}}_c \cdots A_{N}^{\alpha_{N-1} d_N} ,
\end{multline}
and is also illustrated graphically in Fig.~\ref{fig:mps_classification}. The contraction of the model during the forward pass with this additional tensor is similar to the ones of the single MPS and can be visualized in Fig.~\ref{fig:classification}.
% during the forward process that leads to the computation of the previously cited vector of length $C$.

\begin{figure}[h!]
    \centering
    \begin{tikzpicture}[inner sep=0.5mm]
        \node[tensor, minimum size=2.0em] (1) at (1, 0) {\tiny$A_1$};
        \draw[-] (1) -- (1, -.7);
        \node[] (2) at (2, 0) {$\ldots$};
        % \draw[-] (2) -- (2, -.7);
        \node[tensor, minimum size=2.0em] (3) at (3, 0) {\tiny$A_{\frac{n}{2}}$};
        \draw[-] (3) -- (3, -.7);

        % label tensor
        \node[tensor2, minimum size=2.0em] (c) at (4,0) {$C$};
        \draw[-] (c) -- (4, +0.7);
    
        % index where we are sampling from
        \node[tensor, inner sep=0pt, minimum size=2.0em] (4) at (4+1, 0) {\tiny$A_{\frac{n}{2}+1}$};
        \draw[-] (4) -- (4+1, -0.7);
        
        \node[] (5) at (5+1, 0) {$\ldots$};
        \node[tensor, minimum size=2.0em] (6) at (6+1, 0) {\tiny$A_n$};
        \draw[-] (6) -- (6+1, -.7);
        
        % horizontal lines
        \foreach \i in {1,...,2} {
            \pgfmathtruncatemacro{\iplusone}{\i + 1};    
            \draw[-] (\i) -- (\iplusone);
        };
        \draw[-] (3) -- (c);
        \draw[-] (c) -- (4);
    
        \foreach \i in {4,...,5} {
            \pgfmathtruncatemacro{\iplusone}{\i + 1};
            \draw[-] (\i) -- (\iplusone);
        };
    
    \end{tikzpicture}
    \caption{This illustrates the MPS architecture used for classification in the case of an additional central tensor. The blue tensors constitute the MPS, while the red component represents the additional tensor that enables multiple label classes for classification.}
    \label{fig:mps_classification}
\end{figure}

\begin{figure}[h!]
\centering
    \begin{tikzpicture}[inner sep=0.5mm]
        \foreach \i in {1,...,2} {
            \node[tensor] (\i) at (\i, 0) {$A_\i$};
            \node[tensor] (x\i) at (\i, -1){};
            \node[font = \footnotesize] at (\i, -1.35) {$\phi(x_\i)$};
            \draw[red, -] (\i) -- (x\i);
        };
        % label tensor
        \node[tensor2] (c) at (3,0) {$C$};
        \draw[-] (c) -- (3, +0.7);
        \foreach \i in {3,...,4} {
            \node[tensor] (\i) at (\i+1, 0) {$A_\i$};
            \node[tensor] (x\i) at (\i+1, -1){};
            \node[font = \footnotesize] at (\i+1, -1.35) {$\phi(x_\i)$};
            \draw[red, -] (\i) -- (x\i);
        };
        % horizontal lines
        \foreach \i in {1} {
            \pgfmathtruncatemacro{\iplusone}{\i + 1};
            \draw[-] (\i) -- (\iplusone);
        };
        \draw[-] (2) -- (c);
        \draw[-] (c) -- (3);
        \foreach \i in {3} {
            \pgfmathtruncatemacro{\iplusone}{\i + 1};
            \draw[-] (\i) -- (\iplusone);
        };
    \end{tikzpicture}
    \linebreak
    \begin{tikzpicture}[inner sep=0.5mm, minimum size = 2.0em]
        \node[tensor] (1) at (1, 0) {$L$};
        \node[tensor] (2) at (2, 0) {$A_2$};
        % label tensor
        \node[tensor2] (c) at (3,0) {$C$};
        \draw[-] (c) -- (3, +0.7);
        \node[tensor] (3) at (4, 0) {$A_3$};
        \node[tensor] (4) at (5, 0) {$R$};
        % horizontal lines
        \draw[red, -] (1) -- (2);
        \draw[-] (2) -- (c);
        \draw[-] (c) -- (3);
        \draw[red, -] (3) -- (4);
    \end{tikzpicture}
    \linebreak
    \linebreak 
    \begin{tikzpicture}[inner sep=0.5mm, minimum size = 2.0em]
        \node[tensor] (1) at (1, 0) {$L$};
        % label tensor
        \node[tensor2] (c) at (2,0) {$C$};
        \draw[-] (c) -- (2, +0.7);
        \node[tensor] (2) at (3, 0) {$R$};        
        \draw[red, -] (1) -- (c);
        \draw[red, -] (c) -- (2);
    \end{tikzpicture}
    \linebreak 
    \linebreak 
    \begin{tikzpicture}[inner sep=0.5mm, minimum size = 2.0em]
        % output tensor
        \node[tensor2] (c) at (1,0){$y$};
        \draw[-] (c) -- (1, +0.7);
    \end{tikzpicture}
        
    \caption{This illustrates an example of how MPS and inputs are contracted in the forward pass of the classification, considering the case $N=4$. The red lines indicate the indices that are being contracted in each step.}
    \label{fig:classification}
\end{figure}

An ensemble of MPSs is fundamentally equivalent to the single MPS with the additional central tensors. Each matrix in the single MPS must be block-diagonal with bond dimension $d*C$ and blocks of size $d$, where each of those blocks corresponds to a different MPS from the ensemble. The primary advantage of using a single model with an additional tensor is the capability to store and compute everything with just one MPS. This contrasts with the ensemble approach, which requires multiple MPSs. For instance, the MPS with a central tensor can be effective for low-dimensional inputs with a low bond dimension. However, in cases involving higher-dimensional data with a higher bond dimension, using an ensemble of MPSs can lead to an easier path to optimizing the MPS model, which is desirable in our case instead of the single model with a central tensor used in multiple other studies~\cite{peps:supervised}~\cite{mps:supervised}.

In both scenarios, the final result after contraction is a vector of length $C$, where the squared entries of $C$ are proportional to the probability for the input to belong to each given class, i.e., $p(c=i|\mathbf{x}) = \frac{1}{Z} \cdot y_i^2$, with $Z = \sum_c y_c^2$ being a normalization constant and $y$ being the output of the MPSs.
% the final result after the contractions of the MPSs with the embedded input.

In contrast to the typical classification settings, our model's output does not conform to the equation $\sum_{i=1}^{C} Y_i = 1 $, where $Y_i$ is the probability of the class $i$. This is because our model does not directly compute $p(c = i|\mathbf{x})$. Instead, our classification is determined (in the case of an ensemble of MPSs) by comparing which of the MPSs results in a higher squared value after the contraction with the input.
% It is important to note that a softmax activation function is commonly used in classification tasks to obtain a probabilistic interpretation of the output, but we intentionally avoid it in our case.
We avoid the commonly used softmax activation function because we aim to generate new samples, and the usual probabilistic constraint set using a softmax activation function would alter the capabilities of the MPS for generating purposes, described later in Sect.~\ref{subsec:generation}.

% , the fact of not normalizing the MPSs, allows us to also make them work on datasets with unbalanced classes, where the norm of the tensor networks will be proportional to the weight of the corresponding class. 
We adopt the following initialization scheme for our Matrix Product State (MPS):

\begin{equation}
\label{eq:initialization}
  \text{mps}[i, :, :, j] = \frac{I_D}{\sqrt{d}}, \quad \forall j \in \{0, \ldots, d-1 \}, i \in \{1, \ldots, N\} .
\end{equation}

This ensures that each sequential operation on the preceding matrix, i.e., the multiplication of each matrix with the vectors \( R \) and \( L \) outlined in Algorithm~\ref{algo:classification}, behaves as an identity matrix.

\begin{algorithm}[H]
\caption{MPS Classification}
\label{algo:classification}
\begin{algorithmic}[1]
\Procedure{classification}{mps, C, input}
  \State $mps_{nlr} \leftarrow \sum_{n,e} mps_{nlre} input_{ne}$
  \State $L \leftarrow mps[0, 0]$
  \State $R \leftarrow mps[-1, :, 0]$
  \For{$i$ \textbf{in} range(1, N//2)}
    % \State $L \leftarrow einsum('r,rl\rightarrow l', L, mps[i])$
    \State $L \leftarrow  \sum_{r}L_r mps[i]_{rl}$

    % \State $R \leftarrow einsum('l,rl\rightarrow r', R, mps[-i])$
    \State $R \leftarrow \sum_{l} R_l mps[-i]_{rl}$

  \EndFor
  \State $y \leftarrow \sum_{r, l} L_r C_{crl} R_l$
  \State return $y^2$
\EndProcedure
\end{algorithmic}
\end{algorithm}

In Algorithm~\ref{algo:classification}, we use the following index conventions:
\begin{itemize}
    \item $n$ is the site index, referring to a specific site in the MPS representation. It corresponds to the blue circles in the tensor network diagram and has size \( N \), which is equal to the input dimension.
    \item $l$ and $r$ are bond indices that connect different sites. These correspond to the horizontal lines in the TN diagram, representing the outgoing bonds between adjacent sites. They have size \( D \), which is the bond dimension.
    \item $e$ is the embedding index, which connects each site to the embedded input \( \phi(x) \). This index encodes the data fed into the MPS and has size \( d \), which is the physical dimension.
    \item $c$ is the class index. A separate MPS is assigned to each class \( c \), as represented in the final summation.
\end{itemize}
These index definitions clarify how the tensor contractions are performed during the forward pass in the MPS classification procedure, using the Einstein summation notation.

Although past research ~\cite{hrinchuk2019tensorized} has explored non-trivial initialization configurations, we opt for this simpler and more intuitive approach which encourages stability during training.
% In this scheme, the initialization of the network, which is described in Eq.~\ref{eq:initialization}, is set as non-trainable, and an additional trainable parameter is introduced.
The rank-3 tensors present at each network site assume the form \( A_s = \frac{I_d}{\sqrt{d}} + \hat{A}^s \), where only \( \hat{A}^s \) is trainable. Each entry of \( \hat{A}^s \) is initialized from a normal distribution \( \hat{A}^s \sim \mathcal{N}(0,\,\sigma^{2}) \), where \( \sigma \) requires manual adjustment. The introduction of noise serves to disrupt symmetries and prevent convergence to local minima during the initial stages.
Adding weight decay to the model parameters will only affect the matrices $\{\hat{A}^s\}_s$ and will have no effect on the initial settings except to remove the noise, leading to a more stable system during training while using weight decay.

During training, we minimize the cross-entropy loss in our classification settings with the ensemble of MPSs, which is equivalent to minimizing the Kullback–Leibler (KL) divergence between a single label model and the data coming from that class, as described in~\cite{mps:samples}.
% This entire training procedure can be performed using PyTorch's automatic differentiation, which plays a crucial role during the gradient descent phase~\cite{gradient_descent} in training, performed with Adam optimizer~\cite{adam}.

With this method, we avoid using the DMRG-based method during the optimization of the MPS in~\cite{mps:supervised}. Our approach carries the drawback that it could result in a non-normalized MPS; however, our method addresses this and effectively samples from such non-normalized distributions.
% As mentioned earlier, our approach carries the drawback that it could result in a non-normalized Matrix Product State (MPS). However, as elucidated in Sect.~\ref{subsec:generation}, this apparent limitation, particularly noticeable during the sampling phase, is effectively addressed. Throughout our study, we will encounter instances of non-normalized MPS.
% It's noteworthy that all MPS differing solely by a norm define the same PDF, simplifying our understanding of their role in generative processes.

\subsubsection{Generation with MPS}
\label{subsec:generation}

While previous works have employed Metropolis-Hastings and similar Markov chain Monte Carlo (MCMC) methods for sampling from MPS models~\cite{mps:unsupervised}\cite{mps:discrete}\cite{mps:unbiased_mcmc}, we perform an exact sampling approach. An iterative method is used to sample coordinate after coordinate from a one-dimensional non-normalized PDF. A noise vector $\nu$ is used as the input of our generative model, where each component $\nu_i$ of the noise vector will correspond to the quantile of our sample $\hat{x_i}$, i.e.,
\begin{equation}
p(x_i \leq \hat{x_i} \mid x_{1}, \dots, x_{i-1}) = \nu_i.
\end{equation}
This approach eliminates the need for MCMC methods, which rely on constructing a Markov chain to sample from complex distributions.  Traditional MCMC algorithms, such as Metropolis-Hastings, require long iterative stochastic sampling processes to ensure convergence and representative sampling. In contrast, our method is deterministic and directly samples from the desired distribution. Without relying on the random fluctuations of MCMC, this approach avoids issues like autocorrelation, burn-in periods, and convergence diagnostics.

Furthermore, the exact nature of our sampling approach prevents biases that arise in MCMC due to its random walk behavior, which can lead to inefficiencies and inaccuracies. By generating samples directly, our method provides a more efficient and reliable alternative for sampling from the MPS.

% This approach eliminates the need for MCMC methods, which rely on constructing a Markov chain to sample from complex distributions.  MCMC methods, such as Metropolis-Hastings, involve iterative, stochastic sampling processes that require long chains to ensure convergence and representative sampling. In contrast, our method is deterministic and directly samples from the desired distribution without relying on the random fluctuations of MCMC. This eliminates issues associated with MCMC, such as autocorrelation, burn-in periods, and the need for convergence diagnostics.

% Additionally, the exact nature of our sampling approach avoids bias that can arise in MCMC methods. Specifically, MCMC chains can suffer from bias due to the random walk nature of the sampling process, leading to inefficiencies and potential inaccuracies. Our method directly generates samples without these biases, offering a more efficient and reliable approach for sampling from the MPS.

Once the ensemble of MPSs has been trained for classification, its unique structure allows us to generate new samples from it. This is facilitated by the fact that any probability distribution over multiple variables can be expressed as follows, using the chain rule for probabilities:
\begin{equation}
    p(x_1, x_2, ..., x_n)
= \prod_{i = 1}^n p(x_i|\{x_j\}_{j<i}) ,
\end{equation}
and due to the structure of the tensor network and the choice of embedding function, the conditional distributions $p(x_i|\{x_j\}_{j<i})$ can be computed using the reduced density matrix. The sampling procedure is outlined in the following pseudocode for Algorithm~\ref{algo:sample}:

\begin{algorithm}[H]
\caption{MPS Sampling}
\label{algo:sample}
\begin{algorithmic}[1]
\Procedure{sample}{$\text{mps}, \nu$}
\State $samples \leftarrow \text{zeros}(N)$
\For{$i$ \textbf{in} \textbf{range}$(N)$}
\State $V = \text{density\_matrix}(i, \text{mps}, \text{samples})$
\State $\text{pdf}(x) = \phi(x) \cdot V \cdot \phi(x)$
\State $\text{cdf}(x) = \int_{-1}^x \text{pdf}(y) \, dy$
\State $samples[i] = \text{cdf}^{-1}(\nu[i])$
\EndFor
\State \textbf{return} $samples$
\EndProcedure
\end{algorithmic}
\end{algorithm}
Fig.~\ref{fig:density_matrix} shows the reduced density matrix produced at step $i$ by Algorithm~\ref{algo:sample}.
\begin{figure}[h!]
    \centering
    \begin{tikzpicture}[inner sep=0.5mm]

        \node[tensor, minimum size=2.0em] (1) at (1, 0) {\tiny$A_1$};
        \node[tensor, minimum size=2.0em] (1a) at (1, -2.7) {\tiny$A_1$};

        \node[tensor](x1) at (1, -0.85) {};
        \node[tensor](x1a) at (1, -1.85) {};
        \node[font = \footnotesize] at (1, -1.1) {$\phi(x_1)$};
        \node[font = \footnotesize] at (1, -1.6) {$\phi(x_1)$};

        \node[font=\footnotesize] (2) at (2, 0) {$\ldots$};
        \node[font=\footnotesize] (2a) at (2, -2.7) {$\ldots$};

        \node[tensor, minimum size=2.0em] (3) at (3, 0) {\tiny$A_{i-1}$};
        \node[tensor, minimum size=2.0em] (3a) at (3, -2.7) {\tiny$A_{i-1}$};
        
        \node[tensor](x3) at (3, -0.85) {};
        \node[tensor](x3a) at (3, -1.85) {};
        \node[font = \footnotesize] at (3, -1.1) {$\phi(x_{i-1})$};
        \node[font = \footnotesize] at (3, -1.6) {$\phi(x_{i-1})$};

        \node[tensor, minimum size=2.0em] (4) at (4,0) {\tiny$A_i$};
        \node[tensor, minimum size=2.0em] (4a) at (4, -2.7) {\tiny$A_i$};
        \draw[-] (4) -- (4, -0.7);
        \draw[-] (4a) -- (4, -2.0);

        \node[tensor, minimum size=2.0em] (5) at (5, 0) {\tiny$A_{i+1}$};
        \node[tensor, minimum size=2.0em] (5a) at (5, -2.7) {\tiny$A_{i+1}$};

        \draw[-] (5) -- (5a);

        % \node[font = \footnotesize](x5) at (5, -1.1) {$\phi(x_{i+1})$};
        % \node[font = \footnotesize](x5) at (5, -1.6) {$\phi(x_{i+1})$};
        % \node[tensor](x5) at (5, -0.85) {};
        % \node[tensor](x5a) at (5, -1.85) {};
        % \draw[-] (5) -- (x5);
        % \draw[-] (x5a) -- (5a);

        % index where we are sampling from
        \node[font=\footnotesize] (6) at (6, 0) {\tiny$\ldots$};
        \node[font=\footnotesize] (6a) at (6, -2.7) {\tiny$\ldots$};

        \node[tensor, minimum size=2.0em] (7) at (7, 0) {\tiny$A_{n}$};
        \node[tensor, minimum size=2.0em] (7a) at (7, -2.7) {\tiny$A_{n}$};

        \draw[-] (7) -- (7a);

        \draw[-] (1) -- (x1);
        \draw[-] (x1a) -- (1a);

        \draw[-] (3) -- (x3);
        \draw[-] (x3a) -- (3a);

        % horizontal lines
        \foreach \i in {1,...,6} {
            \pgfmathtruncatemacro{\iplusone}{\i + 1};
    
            \draw[-] (\i) -- (\iplusone);
            \draw[-] (\i a) -- (\iplusone a);
        }
        
        % \draw[-] (1.west) .. controls +(-1.5, 1) and +(1.5, 1) .. (5.east);
    \end{tikzpicture}
    \caption{Visual description of the tensor contractions that produce the matrix $V_{i, \{x_j\}_{j<i}}$, used to calculate the conditional probability $p(x_i|x_{1}, ..., {x_{i-1}}) = \phi(x_i)\cdot V_{i, \{x_j\}_{j<i}}\cdot \phi(x_i)$ during the $i$-th iteration of our sampling algorithm.}
    \label{fig:density_matrix}
\end{figure}

% In the case of our training technique, since it can lead to non-normalized MPSs as discussed in Sect.~\ref{subsec:classification}, we will have to deal with non-normalized probability dunsity functions.
The challenge of the non-normalized MPSs is addressed by normalizing the cumulative distribution functions before sampling from them. An alternative approach could be to calculate the norm of the MPS before each sampling step and divide each site by the $n$-th root of the norm, which could be computed using the contractions visualized in Fig.~\ref{fig:norm}, but would require more contraction calculations during training and therefore be less efficient.

\begin{figure}[h!]
\centering
    \begin{tikzpicture}[inner sep=0.5mm]

        \node[tensor] (1) at (1, 0) {$A_1$};
        \node[tensor] (1a) at (1, -1.5) {$A_1$};
        \draw[-] (1) -- (1a);

        \node[font=\footnotesize] (2) at (2, 0) {$\ldots$};
        \node[font=\footnotesize] (2a) at (2, -1.5) {$\ldots$};
    
        % label tensor
        \node[tensor] (3) at (3,0) {$A_i$};
        \node[tensor] (3a) at (3, -1.5) {$A_i$};
        \draw[-] (3) -- (3a);

        \node[font=\footnotesize] (4) at (4, 0) {$\ldots$};
        \node[font=\footnotesize] (4a) at (4, -1.5) {$\ldots$};

        \node[tensor] (5) at (5, 0) {$A_n$};
        \node[tensor] (5a) at (5, -1.5) {$A_n$};
        \draw[-] (5) -- (5a);

        % horizontal lines
        \foreach \i in {1,...,4} {
            \pgfmathtruncatemacro{\iplusone}{\i + 1};
    
            \draw[-] (\i) -- (\iplusone);
            \draw[-] (\i a) -- (\iplusone a);

        }
        
    \end{tikzpicture}
        
    \caption{How to compute the norm of an MPS, using Penrose graphical notation.}
    \label{fig:norm}
\end{figure}

% ??? It's important to note that this approach is primarily suitable for MPS with a small 'n' because computing the matrices $V_i$ becomes impractical if we use the approach described in Chapter~\ref{subsec:classification}, which involves dividing each vector obtained during the MPS contraction.

% More precisely, during the computation of the cumulative probability density and the sampling phase, we will have to make sure to make the procedure differentiable in order to apply the methods described in Sect.~\ref{sec:tensorgan}, since we will need to calculate the derivative with respect to the network's parameters in the gradient descend phase of the GAN-style training.
The computation of the cumulative probability density and sampling phase must be differentiable in order to apply the methods described in Sect.~\ref{sec:tensorgan} to facilitate gradient descent during training. If we denote the cumulative distribution function as,
\begin{equation}
    f_{V_{i, \{x_j\}_{j<i}}}(x) \coloneqq \int_{-1}^x \phi(y) \cdot V_{i, \{x_j\}_{j<i}} \cdot \phi(y) \, dy ,
\end{equation}
we need to compute:
\begin{equation}
\label{eq:sample}
    sample(\nu, V_{i, \{x_{j}\}_{j<i}}) = f_{V_{i, \{x_{j}\}_{j<i}}}^{-1}(\nu)
\end{equation}
to obtain a sample given the quantile $\nu$ and the reduced density matrix $V_{i, \{x_j\}_{j<i}}$.

% In this scenario, differentiating the equation $f^{-1}(f(x)) = x$ yields:

% \begin{equation}
%     f'(f^{-1}(x))f'^{-1}(x) = 1
% \end{equation}

% which leads to the following differential equation:

% \begin{equation}
% \label{diff_eq}
% f'(x) = \frac{1}{\text{emb}(x) \cdot V \cdot \text{emb}(x)}
% \end{equation}

% \begin{equation}
% I(x_k) = \int_{-1}^{x_k} \text{emb}(x) \cdot V \cdot \text{emb}(x) \, dx
% \end{equation}

Depending on the choice of $\phi(x)$, there may not exist an explicit solution for $f^{-1}(x)$. Hence, the integral is approximated as a finite sum by dividing the input's support into 1000 bins:

\begin{multline} 
I(x_k) = \int_{-1}^{x_k} \phi(x) \cdot V \cdot \phi(x) \, dx = \\
\sum_{i=0}^k \phi(x_i) \cdot V \cdot \phi(x_i)\Delta x ,
\end{multline}
 where the bin width is denoted as $\Delta x = \frac{1}{1000}$. We define a set of points as $x_k = k \Delta x$ for $k = 0, 1, 2, \ldots, 1000$ in the case of an embedding function with $\text{Supp}(\Phi)=[0,1]$. Then, for a given $\nu$ within the cumulative density function range, we can approximate the inverse function as follows, opting for linear interpolation between the two closest points:

\begin{equation}
f^{-1}(\nu) \approx x_k + \frac{\nu - I(x_k)}{I(x_{k+1}) - I(x_k)} \cdot \Delta x ,
\end{equation}
where $x_k \leq f^{-1}(\nu) \leq x_{k+1}$.
Using this approximation, we can effectively compute the inverse function of the CDF and use it for sampling in Algorithm~\ref{algo:sample}. We generate the vector entries iteratively by sampling points and calculating the conditional reduced density matrix for the next coordinate.  For further details on the impact of binning resolution on accuracy and computational efficiency, refer to \hyperref[app]{Appendix}.

\subsubsection{Avoiding vanishing/exploding values in the contractions}
\label{subsec:normalization}
If we want to initialize our MPS as a stable system, because of the sequential nature of the MPS contraction phase, each of the contractions must be identity operators on the previous tensor; if we initialize each site of the MPS as stated in Eq.~\ref{eq:initialization}, the embedding function must then follow the property:
\begin{equation}
    \label{eq:sumnorm}
 \forall x : \sum_k emb(x)_k = \sqrt{d} .
\end{equation}
However, this condition, in addition to Eq.~\ref{eq:emb}, would limit the choice for the embedding functions excluding those described in Sect.~\ref{subsec:embedding}, so we decided to opt for an alternative approach.

Given that our model can be represented as a linear model, with the exception of the embedding phase of our data, any subsequent scalar multiplication will not have any influence on the classification and final sampling outcomes. This holds true both during the forward pass of the MPS, as well as in instances involving contraction for the derivation of the density matrix $V_{i, \{x_j\}_{j<i}}$. 

To solve the problem of exploding and vanishing values, at each site of the MPS, we divide the vectors of the contraction, i.e., the variables denoted by $L$ and $R$ in Algo.~\ref{algo:classification}, by the value of their largest component after each step. In this way, we will always obtain vectors with a norm contained in the interval $[1, \sqrt{d}]$, avoiding both vanishing and exploding values. 

% ??? Since this procedure is happening during training, the model will learn itself how to deal with this procedure, so that it will not affect the classification performances of it.

A similar technique is implemented during the computation of the reduced density matrix. Specifically, the matrix is divided by the absolute value of its maximum entry,
%while preserving the integrity of the results, 
since the non-normalized PDF defined by Eq.~\ref{eq:pdf} will not be affected by any positive scalar multiplication, allowing us to use the sampling technique in Sect.~\ref{subsec:generation} even after rescaling the matrices.

%TODO: correct text

\subsection{MPS for simultaneous classification and generation}
\label{sec:tensorgan}
As discussed in Sect.~\ref{subsec:generation}, an MPS trained for classification can be used as a generator to emulate the data that was fed during training. However, this procedure would produce a lot of outliers that does not accurately reproduce the original data~\cite{mps:positive_unlabeled_learning}. A solution proposed in~\cite{mps:positive_unlabeled_learning} is to accept generated samples only if the MPS itself outputs a high enough probability that the sample is in the correct class; otherwise, if the output is lower than a pre-determined threshold, the sample is rejected. This removes most outliers in a post-processing step but relies on a manually tuned threshold.
% but the threshold has to be set manually, and this method does not really change the abilities of the network.

We opt to address this challenge directly from the network architecture to diminish the generation of outliers by using the MPS as a generator in a GAN-style setting and letting it compete with a discriminator. There are few regularization techniques for MPS models, for example, weight decay techniques are often used~\cite{mps:supervised}, but dropout can not be used easily as in fully connected networks. This GAN-style setting can also be seen as a regularization method of the MPS training. As seen in Sect.~\ref{sec:result}, our GAN-style training produces more realistic samples, without changing the classification accuracy of the model. Here is an overview of our, MPS-GAN, training process:

\begin{enumerate}
    \item Pre-training the MPS: Before the adversarial training, the MPS can be pre-trained using a classification-oriented approach to obtain realistic samples. We empirically observe that this step helps the MPS learn the underlying patterns. Note that it is an intensive training, up to acceptable classification accuracy for the model, since we will use the score of this pre-trained model as a baseline for the adversarial-trained model. This is necessary for the classification accuracy of the final generator to be superior to the pretrained MPS.
    
    \item Pre-training the discriminator: The discriminator, which can be a fully connected neural network or a specialized network (such as convolutional neural networks in the case of images), is pre-trained on samples generated by the previously trained MPS and the original dataset.
    
    \item Adversarial training: The adversarial training iteratively optimizes both the MPS generator and the discriminator.
    \begin{itemize}
        \item Generator optimization: Generate samples using the MPS, and then optimize it to minimize the discriminator's ability to distinguish between real and generated samples. 
        \item Discriminator optimization: The discriminator is optimized using the standard GAN objective, where it maximizes the probability of assigning the correct label to real samples and the incorrect label to generated samples. 
    \end{itemize}
    \item Classification accuracy check: the classification accuracy is checked every epoch to ensure it remains above the initial classification accuracy threshold. If ever it falls below this threshold, the model is retrained in the classical classification setting. This step ensures that the generator maintains its classification capabilities.
\end{enumerate}

An ensemble of discriminators for the different labels is used to reduce the number of outliers among the sample points, while also helping to distinguish between classes. It also prevents samples from being wrongly identified as real samples if they are unlabeled and assigned to the false class.

\section{Results and discussion}
\label{sec:result}

The results section will primarily concentrate on the generative performance of our method, as our primary objective is to enhance the network's generative capabilities. During the following experiments, we will search for the best MPS model, according to its classification accuracy, 

We demonstrate the generation capabilities of the GAN-style MPS and present an analysis of the latent space. The parameters are set as $D \in \{4,10,30\}, d \in \{4,10,30\}$, with an initial standard deviation of $\sigma = 0.1$ and an initial learning rate of $lr = 0.01$. Additionally, learning rate decay and early stopping procedures are applied.

The bond dimension $D$ plays a crucial role in determining the expressive power of MPS models, as it controls both the number of parameters and the ability to model correlations between variables~\cite{navascues2018bond}. To assess the impact of $D$ on classification performance, an experiment was conducted in which the model was trained across a range of bond dimensions. The results, detailed in \hyperref[app]{Appendix}, indicate that beyond a threshold of $D \approx 4$, classification accuracy does not significantly improve. This suggests that, for low-dimensional datasets, large bond dimensions are not necessary for capturing meaningful correlations. Higher values of $D$ help model long-range dependencies, especially in images~\cite{mps:unsupervised}, but these are absent in our low-dimensional data.

\subsection{GAN-style training}
\label{subsec:result_gan}

In this section, the generative performances of the MPS models are analyzed pre- and post-GAN-style training, with comparisons between Fourier and Legendre embedding functions. The results are based on simulated datasets, the dual 2D Spiral, the 2 Moon, and the Iris datasets, and more details of the simulated data can be found in the \hyperref[app]{Appendix}.

An FID-like score is used to compare the generated samples before and after training the MPS in a GAN-style setting. For a training dataset ${x_1, \ldots, x_n}$ and a set of generated samples ${x_1^{(g)}, \ldots, x_m^{(g)}}$, we denote the respective means as $\hat{\mu}$ and $\hat{\mu}^{(g)}$, and the covariance matrices as $\hat{\Sigma}$ and $\hat{\Sigma}^{(g)}$. The FID-like metric is defined as

\begin{equation}
\|\hat{\mu} - \hat{\mu}^{(g)}\|^2 + \text{Tr}(\hat{\Sigma} + \hat{\Sigma}^{(g)} - 2(\hat{\Sigma}\hat{\Sigma}^{(g)})^{1/2}) .
\end{equation}
This metric quantifies the dissimilarity between the real and generated data based on their means and covariance matrices, where a lower value indicates greater similarity between the two datasets.
\begin{table}[!h]
    \centering
    \caption{Comparison of FID-like score (lower is better) on generated samples for Fourier and Legendre embedding functions, pre- and post-GAN-style training.}
    \label{tab:fid}

    \begin{tabular}{@{}l S[table-format=1.2e-1] S[table-format=1.2e-1] S[table-format=1.2e-1]@{}} 
        \toprule
        & \multicolumn{3}{c}{Datasets} \\ 
        \cmidrule(lr){2-4}
        Methods & {2D Spiral} & {2 Moon} & {Iris} \\ 
        \midrule
        Legendre, pre-GAN  & 4.63e-3  & 1.20e-2  & 6.36e-2  \\
        Fourier, pre-GAN   & 3.25e-3  & 8.79e-4  & 1.02e-1  \\
        Legendre, post-GAN & 4.57e-3  & 5.93e-3  & 4.04e-2  \\
        Fourier, post-GAN  & \textbf{\num{1.52e-3}}  & \textbf{\num{3.02e-4}}  & \textbf{\num{1.96e-2}}  \\
        \bottomrule
    \end{tabular}
\end{table}

The results for both Legendre and Fourier embeddings pre- and post-GAN-style training are shown in Table~\ref{tab:fid}. The Fourier embedding after the GAN-style training generated the lowest FID-like score across all datasets.

\begin{figure}[h!]
\centering
\includegraphics[width=.15\textwidth]{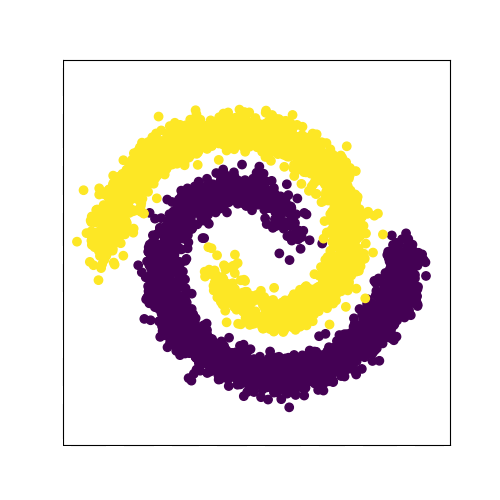}
\includegraphics[width=.15\textwidth]{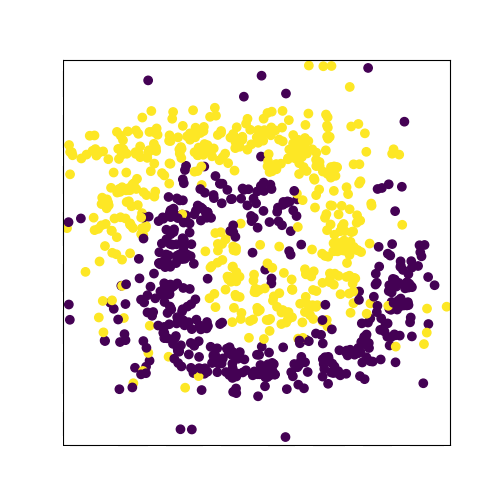}
\includegraphics[width=.15\textwidth]{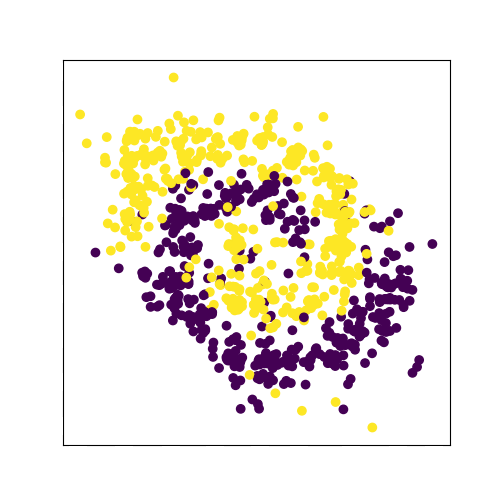}
\caption{Visualization of the 2D Spiral dataset and MPS-generated samples. Left: original dataset. Middle: Samples generated by a classically trained
MPS, optimized using cross-entropy loss. Right: Samples generated by an adversarially trained MPS. Fourier embedding was used.}
\label{spiral}
\end{figure}

\begin{figure}[h!]
\centering
\includegraphics[width=.15\textwidth]{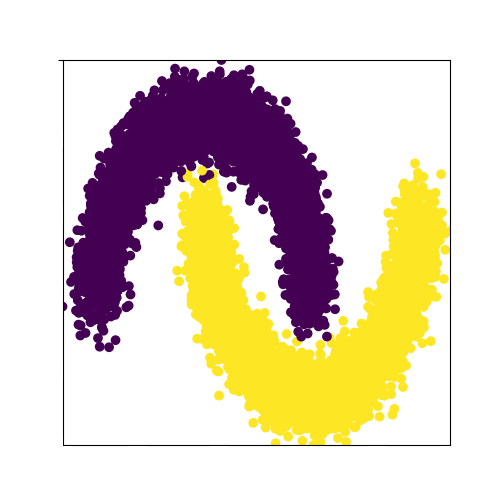}
\includegraphics[width=.15\textwidth]{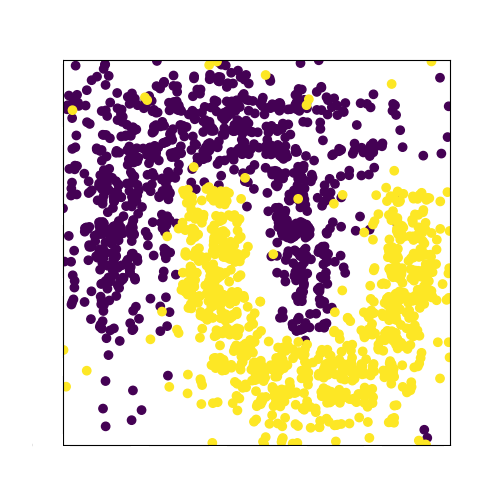}
\includegraphics[width=.15\textwidth]{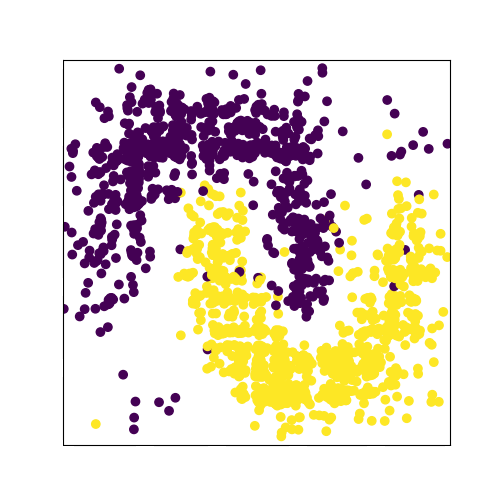}
\caption{Comparison of generated samples on the Two Moons dataset.  Left: The original dataset. Middle: Samples generated by a classically trained MPS, optimized using cross-entropy loss. Right: Samples generated by an adversarially trained MPS. Fourier embedding was used. }
\label{2moon}
\end{figure}

\begin{figure}[h!]
\centering
\includegraphics[width=.15\textwidth]{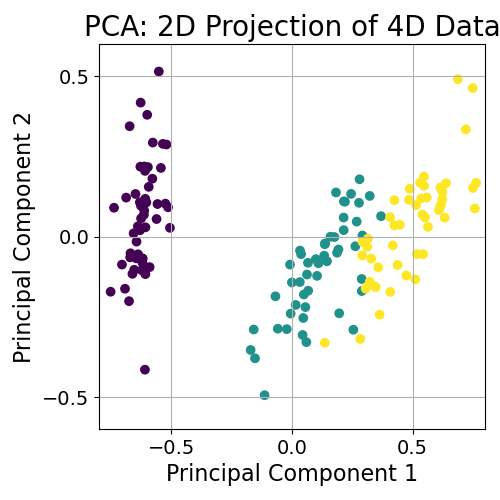}
\includegraphics[width=.15\textwidth]{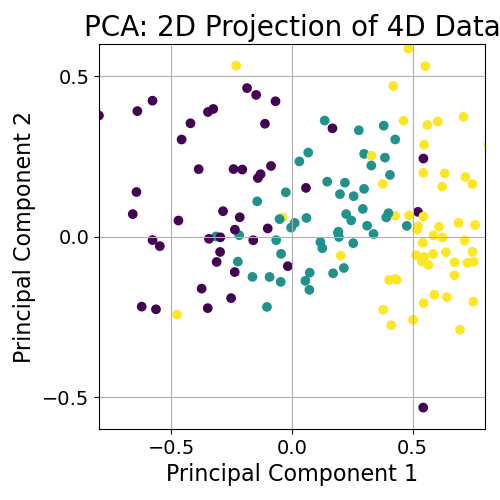}
\includegraphics[width=.15\textwidth]{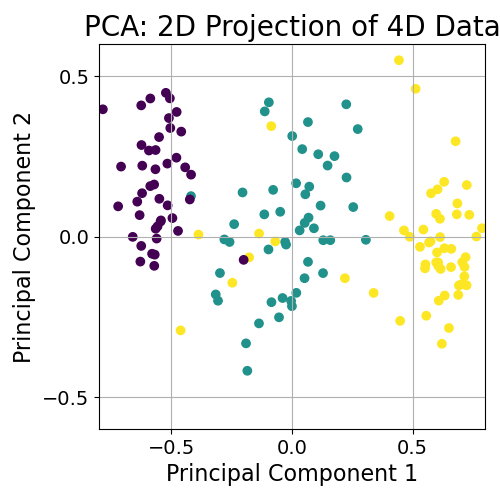}
\caption{2D PCA visualization of generated samples from the Iris dataset. Left: PCA projection of the original dataset. Middle: PCA projection of samples generated by a classically trained MPS model, optimized using cross-entropy loss. Right: PCA projection of samples generated by an adversarially trained MPS model. Fourier embedding was applied to the data.}
\label{iris}
\end{figure}

Fig.~\ref{spiral}, ~\ref{2moon}, and ~\ref{iris} also visually show how GAN-style training affects the result of the samples generated by the ensemble of MPSs.

The number of outliers generated by the model decreases after the GAN-style training, as shown in Table \ref{tab:outliers} which compares the number of
outliers before and after the GAN-style training and for different choices for the embedding functions. We evaluate the percentage of outlier samples, where a sample point is considered an outlier if the average of its k-nearest neighbors in the training data is higher than the maximum k-distance of the original dataset.

\begin{table}[!h]
    \centering
    \caption{Comparison of outlier percentages of generated samples for Fourier and Legendre embedding functions, before and after GAN-style training}
    \label{tab:outliers}
    \begin{tabular}{@{}l S[table-format=1.2e-1] S[table-format=1.2e-1] S[table-format=1.2e-1]@{}} 
        \toprule
        & \multicolumn{3}{c}{Datasets} \\ 
        \cmidrule{2-4}
        Methods & {2D Spiral} & {2 Moon} & {Iris}\\ 
        \midrule
        Legendre, pre-GAN & \num{9.03e-2} & \num{1.00e-1} & \num{3.60e-1} \\
        Fourier, pre-GAN & \num{8.90e-2} & \num{1.13e-1} & \num{4.62e-1} \\
        Legendre, post-GAN & \num{7.63e-2} & \num{1.00e-1} & \textbf{\num{2.20e-1}} \\
        Fourier, post-GAN & \textbf{\num{6.32e-2}} & \textbf{\num{3.40e-2}} & \num{2.36e-1} \\
        \bottomrule
    \end{tabular}
\end{table}

% \begin{table}[h]
%     \centering
%     \caption{Comparison of outlier percentages of generated samples for Fourier and Legendre embedding functions, before and after GAN-style training}
%     \label{tab:outliers}
%     \begin{tabular}{lSSS} 
%         \toprule
%         & \multicolumn{3}{c}{Datasets} \\ 
%         \cmidrule{2-4}
%         Methods & {2D Spiral} & {2 Moon} & {Iris}\\ 
%         \midrule
%         Legendre, pre-GAN & \num{0.09035} & \num{0.10} & \num{0.360} \\
%         Fourier, pre-GAN & \num{0.0890} & \num{0.1130} & \num{0.462} \\
%         Legendre, post-GAN & \num{0.07635} & \num{0.10} & \num{0.220} \\
%         Fourier, post-GAN & \num{0.0632} & \num{0.0340} & \num{0.236} \\
%         \bottomrule
%     \end{tabular}
% \end{table}

The values reported in Table~\ref{tab:outliers} confirm the hypothesis suggested visually by Fig.~\ref{spiral}, ~\ref{2moon}, and ~\ref{iris}, where the number of outliers decreases thanks to the GAN-style training. Additionally, the performances of the models that use Fourier embedding functions have lower FID-like scores across the board compared to those resulting from the Legendre embeddings.

\subsection{Latent space analysis}
\label{subsec:latent}
Within our experimental framework, the latent space dimensionality of our generative model, which follows GAN principles, corresponds to the dataset’s dimensionality. The latent space is equal to $[0,1]^n$, and as described in Sect.~\ref{subsec:generation}, the noise vector $\nu \in [0,1]^n$ part of our latent space satisfies the condition:
\begin{equation*}
p(x_i \leq \hat{x_i} \mid x_{1}, \dots, x_{i-1}) = \nu_i.
\end{equation*}

For an ideally trained model, every point within this latent space corresponds to a realistic sample. Additionally, the set of points in the latent space that do not correspond to realistic samples should have measure of 0, ensuring that the vast majority of the latent space is occupied by realistic data points. Consequently, conventional latent space analyses, such as clustering, become inapplicable, as the latent space does not exhibit well-defined clusters but instead reflects the smooth, continuous distribution of realistic data points.

In scenarios where the probability density function (PDF) is strictly positive and is modeled using the Matrix Product State (MPS) framework, the latent space can be mathematically represented as the hypercube $[0,1]^n$.  While the latent space’s dimensionality is fixed at $\mathbb{R}^n$, independent of the model’s parameters, its structure may be influenced by factors such as the bond dimension. However, as shown in the experiments in \hyperref[app]{Appendix}, the bond dimension has minimal impact in our setting due to the low dimensionality of our datasets. To better understand its influence in more complex scenarios, further experiments with larger datasets would be necessary. Moreover, there exists a one-to-one mapping i.e a bijection, between this latent space and the physical data space. As a result, any topological analysis within the latent space yields trivial findings, since the latent space directly corresponds to the physical space of the data. 

\begin{figure}[h!]
    \centering
    \includegraphics[width=.24\textwidth]{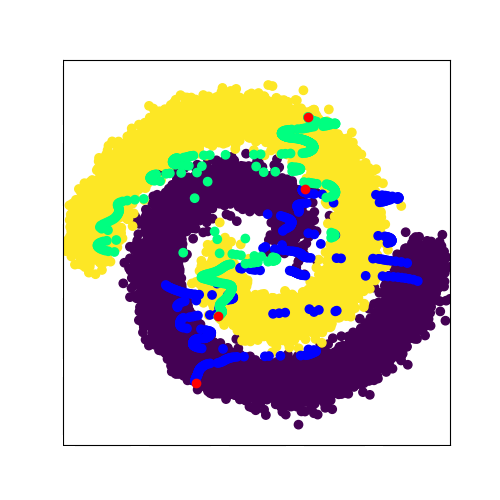}
    \includegraphics[width=.24\textwidth]{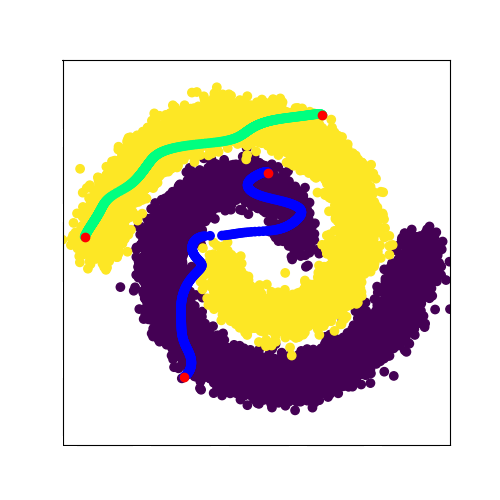}
    \caption{In the background, the original dataset, and in green and blue, the trajectories of samples for, (left) Fourier and (right) Legendre embedding functions, corresponding to the yellow and violet class respectively. The trajectories are obtained by linearly interpolating two points of the latent space of the MPS, highlighted in red.}
    \label{fig:interpolation}
\end{figure}

However, latent space analysis can be performed thanks to the utilization of the MPS architecture in our experimental setup which is itself an ensemble of models, with each model dedicated to a distinct class. This architecture enables us to perform interpolation and comparison exclusively between intra-class samples. A visual representation of this process is presented in Fig.~\ref{fig:interpolation}, where two instances of linear interpolation within the latent space are depicted, one for each class, employing different Fourier and Legendre embedding functions. 

% \begin{figure}[h!]
%     \centering
%     \includegraphics[width=.5\textwidth]{images/interpolation.png}
%     \caption{Trajectories of samples obtained by linearly interpolating between points in the latent space of the MPS used as generator. The red points represent the extremes of the trajectories, and we sampled 500 points for each trajectory. Emb = legendre}
%     \label{fig:interpolation}
% \end{figure}

The sampled trajectories for Legendre-embedded data form a more regular shape, while those for Fourier embeddings appear more chaotic. This structural distinction aligns with their contrasting classification behaviors under input perturbations (see Sect.~\ref{subsec:robustness}).  
\begin{itemize}
    \item Legendre embeddings produce smooth sample trajectories in latent space, reflecting their stable polynomial basis. This regularity leads to well-defined decision boundaries in noiseless conditions, enabling higher initial classification accuracy.
    \item Fourier embeddings, with their oscillatory basis, generate irregular trajectories. While this introduces instability in noise-free settings, the frequency-rich representation improves adaptability to input variations, as shown in Fig.~\ref{fig:perturbation}.
\end{itemize}

In both cases, most samples remain within their original class (Fig.~\ref{fig:interpolation}), demonstrating the MPS’s ability to preserve semantic consistency. However, exceptions occur in low-probability regions where trajectories intersect the "false" class. These intersections result from the MPS’s non-zero probability of sampling across classes—a consequence of its quantum-inspired structure, which avoids strict orthogonality. While undesirable for purity, this property ensures full support over the data manifold, crucial for robust generation under perturbations (see Sect.~\ref{subsec:robustness}).

\subsection{Robustness to perturbations}
\label{subsec:robustness}
We assumed that our model has an input space on $[0,1]^N$ and that the embedding function used in this work transforms the data onto an n-dimensional manifold in a $d\cdot n$ dimensional space. In this section, we consider how adding noise to this manifold can change the performance of the MPS model, depending on the choice of the embedding function. After obtaining a trained MPS, the classification accuracy of the model is observed for increasing values of $\sigma$, which determines the amount of noise $\epsilon$ that is added to the inputs of the MPS after having embedded the data using $\Phi(x)$, with $\epsilon \sim \mathcal{N}(0, \sigma^2)$.

\begin{figure}[h!]
    \centering
    \includegraphics[width=.5\textwidth]{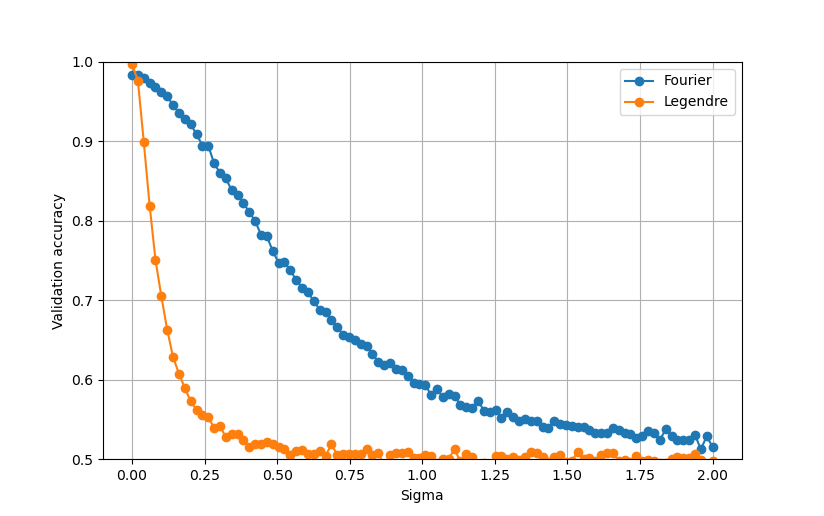}
    \caption{Validation accuracy as a function of $\sigma$, where $\sigma$ represents the standard deviation of the centered normal distribution from which the noise added to the embedded inputs is sampled. In this experiment, the model is trained on clean data and evaluated on noisy inputs, with hyperparameters set to $d = 20$ and $D = 50$. A comparison between Fourier and Legendre embeddings is provided.}
    \label{fig:perturbation}
\end{figure}

Fig.~\ref{fig:perturbation} shows the accuracy performance of two models, one with Fourier and the other with Legendre embedding function, as noise is added to the embedded inputs. 

%The model are trained on the same dataset and use the same set of hyperparameters ($d=30, D = 30, \sigma = 0.1$).

% fourier embeddings: 0.9737 validation accuracy, 100\% training accuracy
% legendre: 0.9972 validation accuracy, 100\% training accuracy
% $d = 30$

As seen in Fig.~\ref{fig:perturbation}, at $\sigma = 0$, the Fourier embedding achieves lower initial classification accuracy compared to the Legendre embedding. A precise theoretical justification remains an open question; however, a potential explanation may be related to the regression properties of these embeddings. 

Legendre embeddings, being based on orthogonal polynomials, appear to be empirically better suited to the dataset, leading to higher validation accuracy. This suggests that the dataset’s underlying patterns can be more naturally captured by polynomial basis functions.

Fourier embeddings introduce oscillatory basis functions, which can lead to high-frequency artifacts in the learned function when trained on limited data, a phenomenon sometimes associated with interpolation instability or Runge’s phenomenon in polynomial interpolation~\cite{trefethen2013approximation}. This effect can cause poor generalization in classification tasks when noise is absent ($\sigma = 0$). In contrast, Legendre embeddings, based on orthogonal polynomials, tend to produce smoother approximations with reduced oscillations, potentially leading to more stable classification boundaries.

However, as noise increases ($\sigma > 0$), Fourier embeddings demonstrate higher robustness, likely due to their ability to capture frequency-based features efficiently. This aligns with prior findings in function approximation theory, where Fourier-based methods excel in encoding structured data but may struggle with instability in noiseless conditions~\cite{tancik2020fourier}. 
%Additional experiments testing the current interpretation can be found in App.~\ref{app:noiseexperiments}.

%Further investigation is required to fully characterize these effects.

\begin{figure}
    \centering
    \includegraphics[width=0.5\textwidth]{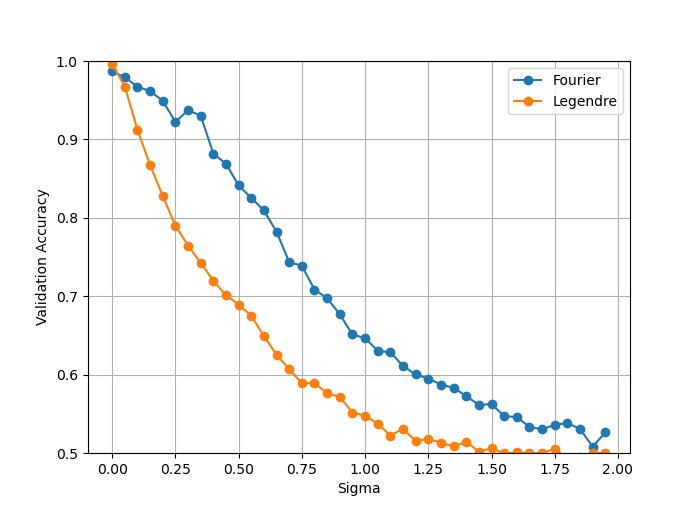}
    \caption{ Validation accuracy as a function of $\sigma$, where $\sigma$ represents the standard deviation of the centered normal distribution from which the noise added to the embedded inputs is sampled. In this experiment, models are trained on noisy data and evaluated on clean inputs, with hyperparameters set to $d = 20$ and $D = 50$. A comparison between Fourier and Legendre embeddings is provided.}
    \label{fig:noisy_training}
\end{figure}

An additional experiment, comparing the classification accuracy of the MPS as $\sigma$ increases, is shown in Fig.~\ref{fig:noisy_training}. In this experiment, noise is introduced during the training phase, whereas in the previous experiment, noise was only added during inference with a fixed trained model.

The results of this experiment align with the current interpretation. While the Legendre embedding achieves better classification performance at $\sigma = 0$, we observe that as noise is introduced during training the Legendre embedding also gains robustness by learning to fit the noise. Conversely, the Fourier embedding retains its previous robustness and performance remains consistent even at $\sigma = 1$, as it is inherently adapted to handle noise without requiring specialized learning.

\section{Conclusion and future work}
\label{sec:conclusion}
In this study, we explore the MPS and its applications in the field of machine learning. We demonstrate the fundamental structure of MPS and its utility in machine learning for tasks related to classification and generative modeling. We expand upon the various canonical forms of MPS and justify a more accessible training approach for its applications in machine learning, specifically avoiding the necessity for the sweeping algorithm.

To facilitate the discussion, we identified and examined the essential criteria for an embedding function that enables accurate sampling from a non-normalized model trained for classification. We demonstrate the computation of the reduced density matrix and establish its positive semi-definiteness. Furthermore, we introduce and contrast the widely adopted Fourier embedding with our novel proposition of using Legendre polynomials. Our empirical findings indicate that the former yields superior generative results.

Building upon principles from GANs, we leverage dual roles of the MPS, allowing it to serve both as a generator and a classifier simultaneously. This enhances training and generative performance without compromising the model's classification accuracy. Notably, this approach results in a reduction of the number of outliers generated during the sampling process, yielding samples with more favorable FID-like scores when compared to classical training techniques for MPSs.

Within the framework of this procedure, we introduce a latent space representation for the MPS model when using it as a generator and subject it to analysis. We investigate the impact of introducing perturbations after data embedding on the classification accuracy for both Fourier and Legendre embeddings, where the former demonstrated greater resilience in the presence of increasing noise.

%The application of Matrix Product States to data sets, particularly those consisting of images, has been impeded by certain technical challenges. However, it is of considerable interest to investigate the applicability of the method delineated in Sect.~\ref{sec:tensorgan} to well-known datasets like MNIST and Fashion MNIST and even combine it with other techniques developed to diminish the number of outliers produced by the MPS, as outlined in~\cite{mps:positive_unlabeled_learning} and briefly presented in Sect.~\ref{sec:tensorgan}. This inquiry seeks to ascertain whether the method can effectively enhance the generation of more realistic samples. Additionally, it is conceivable that the training paradigm inspired by GANs can be seamlessly integrated into other tensor network architectures, such as PEPS. This extension could shed light on whether the reduction in outlier samples, achieved through this training approach, persists in such model configurations.

In further research, alternative embedding functions can be explored. For instance, the use of the Welsh basis, a non-continuous set of orthonormal functions,  warrants exploration. Additionally, investigating non-orthogonal bases is promising, leveraging the research developed in this work on the role of the embedding function in calculating the reduced density matrix. Furthermore, it is interesting to investigate how perturbations
%as introduced in Sect.~\ref{subsec:robustness}, 
may impact the classification accuracy, particularly when various embedding functions are employed. 

While our experiments focus on low-dimensional datasets, future work should explore the impact of bond dimension $D$ on higher-dimensional datasets. In such settings, larger bond dimensions may be necessary to capture long-range correlations and complex dependencies, which are not prominent in our current datasets. Investigating the interplay between bond dimension, dataset dimensionality, and model expressivity would provide valuable insights into the scalability and applicability of MPS-based models to more complex machine learning tasks.

Furthermore, the MPS holds promise for integration with other neural network paradigms, potentially as a latent space representation within a variational autoencoder or a normalizing flow in manifold learning frameworks~\cite{NEURIPS2023_572a6f16}, where the density can be directly estimated mitigating the need of approximate inference.

\section{Acknowledgements}
This project was supported by grants \#2022-531 and \#2022-643 of the Strategic Focus Area “Personalized Health and Related Technologies (PHRT)” of the ETH Domain (Swiss Federal Institutes of Technology). The authors acknowledge Yolanne Yi Ran Lee for the thoughtful discussions and for significant contributions in improving the writing style and clarity of the manuscript.

\newpage

% \printbibliography %Prints bibliography

\bibliographystyle{ieeetr}
\bibliography{biblio_corrected}% \bibliographystyle{unsrt}

\newpage

\appendix
\label{app}

\section*{Penrose tensor notation in detail}\label{app:penrose}
In this section, we provide additional details on the equations used throughout the work. We explicitly present the formulas that were previously expressed only using Penrose tensor notation.
\begin{itemize}
    \item Probability distribution representation, Eq.~\ref{eq:mpspdf}:

\begin{multline*}
p(x_1, \ldots, x_n) = \\
 \left[ \sum_{\{d_i\}} W^{d_1,d_2, ...,d_{n-1},d_n} \Phi(\mathbf{x})^{d_1, d_2, ...,d_{n-1},d_n} \right]^2 = \\
\left[ \begin{tikzpicture}[inner sep=0.5mm, baseline=(current bounding box.center)]
    \foreach \i/\label in {1/x_1,2/x_2,3/\ldots,4/x_{n-1},5/x_n} {
        \ifnum\i=3
        \node[font=\footnotesize] (x\i) at (\i, 0) {$\ldots$};
        \node[font=\footnotesize] (y\i) at (\i, -.5) {};
        \node[font=\footnotesize] (y\i*) at (\i, -1) {};
        \else
        \node[tensor, font=\footnotesize] (x\i) at (\i, 0) {};
        \node[tensor, font=\footnotesize] (y\i) at (\i, -.5) {};
        \draw[-] (x\i) -- (y\i);
        \node[font = \footnotesize] at (\i, -.75) {$\phi(\label)$};
\fi
    }
    \foreach  \i in {1,...,4} {
        \pgfmathtruncatemacro{\next}{\i + 1}
        \pgfmathtruncatemacro{\nextstar}{\i + 1}
        \draw[-] (x\i) -- (x\next);
    }
\end{tikzpicture} \right] ^2 = \\
   \begin{tikzpicture}[inner sep=0.5mm, baseline=(current bounding box.center)]
    \foreach \i/\label in {1/x_1,2/x_2,3/\ldots,4/x_{n-1},5/x_n} {
        \ifnum\i=3
        \node[font=\footnotesize] (x\i) at (\i, 0) {$\ldots$};
        \node[font=\footnotesize] (x\i*) at (\i, -2) {$\ldots$};
        \node[font=\footnotesize] (y\i) at (\i, -.5) {};
        \node[font=\footnotesize] (y\i*) at (\i, -1) {};
        \node[font = \footnotesize] at (\i, -1) {$\label$};
        \else
        \node[tensor, font=\footnotesize] (x\i) at (\i, 0) {};
        \node[tensor, font=\footnotesize] (x\i*) at (\i, -2) {};
        \node[tensor, font=\footnotesize] (y\i) at (\i, -.5) {};
        \node[tensor, font=\footnotesize] (y\i*) at (\i, -1.5) {};
        \draw[-] (x\i) -- (y\i);
        \draw[-] (x\i*) -- (y\i*);
        \node[font = \footnotesize] at (\i, -.75) {$\phi(\label)$};
        \node[font = \footnotesize] at (\i, -1.25) {$\phi(\label)$};
\fi
    }
    \foreach  \i in {1,...,4} {
        \pgfmathtruncatemacro{\next}{\i + 1}
        \pgfmathtruncatemacro{\nextstar}{\i + 1}
        \draw[-] (x\i) -- (x\next);
        \draw[-] (x\i*) -- (x\nextstar*);
    }
\end{tikzpicture}.
\end{multline*}

\item Definition of $B$, Eq.~\ref{eq:bdef}:

\begin{equation*}
    B \coloneqq \int \phi(x_i) \otimes \phi(x_i) dx_i = \int \begin{tikzpicture}[inner sep=0.5mm, baseline={(current bounding box.center)}]
        \node[tensor] (x) at (0, -0.5) {};
        \node[tensor] (x*) at (0, 0.5) {};
        \node[font = \footnotesize] at (0, 0.25) {$\phi(x_i)$};
        \node[font = \footnotesize] at (0, -0.25) {$\phi(x_i)$};
        \draw[-] (0,-0.75) -- (x);
        \draw[-] (0,0.75) -- (x*);
    \end{tikzpicture} dx_i.
\end{equation*}

\item Marginal probability computation, Eq.~\ref{eq:cond_prob1}:
\begin{align*}
    p(x_i) &=  
    \int \ldots \int p(x_1,\ldots, x_n) \,dx_1 \ldots \widehat{dx_i} \ldots dx_n \\
    &= \int \ldots \int
    \begin{tikzpicture}[inner sep=0.5mm, baseline=(current bounding box.center)]
        \foreach \i/\label in {1/x_1,2/\ldots,3/x_n} {
            \ifnum\i=2
                \node[font=\footnotesize] (x\i) at (\i, 1) {$\ldots$};
                \node[font=\footnotesize] (x\i*) at (\i, -1) {$\ldots$};
            \else    
                \node[tensor, font=\footnotesize] (x\i) at (\i, 1) {};
                \node[tensor, font=\footnotesize] (x\i*) at (\i, -1) {};
                \node[tensor, font=\footnotesize] (y\i) at (\i, 0.5) {};
                \node[tensor, font=\footnotesize] (y\i*) at (\i, -0.5) {};
                \draw[-] (x\i) -- (y\i);
                \draw[-] (x\i*) -- (y\i*);
                \node[font = \footnotesize] at (\i, 0.25) {$\phi(\label)$};
                \node[font = \footnotesize] at (\i, -0.25) {$\phi(\label)$};
            \fi
        }
        \foreach  \i in {1,...,2} {
            \pgfmathtruncatemacro{\next}{\i + 1}
            \draw[-] (x\i) -- (x\next);
            \draw[-] (x\i*) -- (x\next*);
        }
    \end{tikzpicture}
    \,dx_1 \ldots \widehat{dx_i} \ldots dx_n \\
    &\makebox[\columnwidth-50px][r]{
    \begin{tikzpicture}[inner sep=0.5mm, baseline=(current bounding box.center)]
        \foreach \i/\label in {1/x_1,2/\ldots,3/x_i,4/\ldots,5/x_n} {
            \ifnum\i=3
                \node[tensor, font=\footnotesize] (x\i) at (\i, 1) {};
                \node[tensor, font=\footnotesize] (x\i*) at (\i, -1) {};
                \node[tensor, font=\footnotesize] (y\i) at (\i, 0.5) {};
                \node[tensor, font=\footnotesize] (y\i*) at (\i, -0.5) {};
                \draw[-] (x\i) -- (y\i);
                \draw[-] (x\i*) -- (y\i*);
                \node[font=\footnotesize] at (\i, 0.25) {$\phi(x_i)$};  
                \node[font=\footnotesize] at (\i, -0.25) {$\phi(x_i)$};       
            \else
                \ifnum\i=2
                    \node[font=\footnotesize] (x\i) at (\i, 1) {$\ldots$};
                    \node[font=\footnotesize] (x\i*) at (\i, -1) {$\ldots$};
                \else
                    \ifnum\i=4
                        \node[font=\footnotesize] (x\i) at (\i, 1) {$\ldots$};
                        \node[font=\footnotesize] (x\i*) at (\i, -1) {$\ldots$};
                    \else
                        \node[tensor, font=\footnotesize] (x\i) at (\i, 1) {};
                        \node[tensor, font=\footnotesize] (B) at (\i, 0) {$B$};
                        \node[tensor, font=\footnotesize] (x\i*) at (\i, -1) {};
                        \draw[-] (x\i) -- (B); 
                        \draw[-] (B) -- (x\i*); 
                    \fi
                \fi
            \fi
        }
        \foreach  \i in {1,...,4} {
            \pgfmathtruncatemacro{\next}{\i + 1}
            \draw[-] (x\i) -- (x\next);
            \draw[-] (x\i*) -- (x\next*);
        }
    \end{tikzpicture}.
    }
\end{align*}
\item Marginal probability computation in case of $B = I_d$, Eq.~\ref{eq:marginalize}

\begin{align*}
p(x_i) =  \sum_{d_1, ..., d_i, d_i', ..., d_n}&W^{d_1, ..., d_i, ..., d_n}\phi(x_i)^{d_i}\\
&W^{d_1, ..., d_i', ..., d_n} \phi(x_i)^{d_i'} = \\
&\begin{tikzpicture}[inner sep=0.5mm, baseline=(current bounding box.center)]
    % \foreach \i/\label in {1/x_1,2/\ldots,3/x_{i-1},4/x_i,5/x_{i+1}, 6/\ldots, 7/x_n} {
    \foreach \i/\label in {1/x_1,2/\ldots,3/x_i, 4/\ldots, 5/x_n} {
        \ifnum\i=3
            \node[tensor, font=\footnotesize] (x\i) at (\i, 1) {};
            \node[tensor, font=\footnotesize] (x\i*) at (\i, -1) {};
            \node[tensor, font=\footnotesize] (y\i) at (\i, 0.5) {};
            \node[tensor, font=\footnotesize] (y\i*) at (\i, -0.5) {};
            \draw[-] (x\i) -- (y\i);
            \draw[-] (x\i*) -- (y\i*);
            \node[font=\footnotesize] at (\i, 0.25) {$\phi(x_i)$};  
            \node[font=\footnotesize] at (\i, -0.25) {$\phi(x_i)$};       
        \else
            \ifnum\i=2
                \node[font=\footnotesize] (x\i) at (\i, 1) {$\ldots$};
                \node[font=\footnotesize] (x\i*) at (\i, -1) {$\ldots$};
                \node[font=\footnotesize] (y\i) at (\i, 0.5) {};
                \node[font=\footnotesize] (y\i*) at (\i, -0.5) {};
                \node[font=\footnotesize] at (\i, 0) {$\label$};
            \else
                \ifnum\i=4
                    \node[font=\footnotesize] (x\i) at (\i, 1) {$\ldots$};
                    \node[font=\footnotesize] (x\i*) at (\i, -1) {$\ldots$};
                    \node[font=\footnotesize] (y\i) at (\i, 0.5) {};
                    \node[font=\footnotesize] (y\i*) at (\i, -0.5) {};
                    \node[font=\footnotesize] at (\i, 0) {$\label$};
                \else
                    \node[tensor, font=\footnotesize] (x\i) at (\i, 1) {};
                    \node[tensor, font=\footnotesize] (x\i*) at (\i, -1) {};
                    \draw[-] (x\i) -- (x\i*); 
                \fi
            \fi
        \fi
    }
    \foreach  \i in {1,...,4} {
        \pgfmathtruncatemacro{\next}{\i + 1}
            \draw[-] (x\i) -- (x\next);
            \draw[-] (x\i*) -- (x\next*);
    }
\end{tikzpicture} .
\end{align*}

\item Definition of $B_i$, Eq.~\ref{eq:b_def}:
\begin{align*}
    B_i \coloneqq \sum_{d_1, ..., d_{i-1}} &W^{d_1,\ldots, d_{i-1}, d_i, \ldots ,d_n} \\
    &\phi(x_1)^d_1 \otimes \cdots \otimes \phi(x_{i-1})^{d_{i-1}} = \\
    \sum_{d_1,..., d_{i-1}} &A_{1}^{\alpha_1, d_1}\phi(x_1)^{d_1}\\
    &\cdots\\
    &A_{i-i}^{\alpha_{i-2} \alpha_{i-1}, d_{i-1}}  \phi(x_{i-1})^{d_{i-1}} \\
    &A_{i}^{\alpha_{i-1} \alpha_{i}, d_{i}}\\
    &A_{i+1}^{\alpha_{i} \alpha_{i+1}, d_{i+1}}\\
    &\cdots \\
    &A_{N}^{\alpha_{N-1}, d_N} = \\
    &\makebox[\columnwidth-110px][r]{\begin{tikzpicture}[inner sep=0.2mm, baseline={(current bounding box.center)}]
    \node[tensor, minimum size=2.0em] (x1) at (1, 0) {\tiny $A_1$};
    \node[tensor, font=\footnotesize] (y1) at (1, -.75) {};
    \draw[-] (x1) -- (y1);
    \node[font=\footnotesize] at (1, -1) {$\phi(x_1)$};  
    \node[font=\footnotesize] (x2) at (2, 0) {$\ldots$};
    \node[font=\footnotesize] (y2) at (2, -.5) {};
    \node[font=\footnotesize] at (2, -.75) {};  
    \node[tensor, font=\footnotesize, minimum size=2.0em] (x3) at (3, 0) {\tiny$A_{i-1}$};
    \node[tensor, font=\footnotesize] (y3) at (3, -.75) {};
    \draw[-] (x3) -- (y3);
    \node[font=\footnotesize] at (3, -1) {$\phi(x_{i-1})$};
    \node[tensor, font=\footnotesize, minimum size=2.0em] (x4) at (4, 0) {\tiny$A_i$};
    \node (y4) at (4, -0.75) {};
    \draw[-] (x4) -- (y4); 
    \node[tensor, font=\footnotesize, minimum size=2.0em] (x5) at (5, 0) {\tiny$A_{i+1}$};
    \node (y5) at (5, -0.75) {};
    \draw[-] (x5) -- (y5); 
    \node[font=\footnotesize] (x6) at (6, 0) {$\ldots$};
    \node[font=\footnotesize] (y6) at (6, -.5) {};
    \node[font=\footnotesize] at (6, -.75) {};
    \node[tensor, font=\footnotesize, minimum size=2.0em] (x7) at (7, 0) {\tiny$A_n$};
    \node (y7) at (7, -0.75) {};
    \draw[-] (x7) -- (y7); 
    % \foreach \i/\label in {1/x_1,2/\ldots,3/x_i,4/\ldots,5/x_n} {   
            % \ifnum\i=3
            %     \node[font=\footnotesize] (x\i) at (\i, 0) {$\ldots$};
            %     \node[font=\footnotesize] (y\i) at (\i, -.5) {};
            %     \node[font=\footnotesize] at (\i, -.75) {$\label$};
            % \else
            %     \node[tensor, font=\footnotesize] (x\i) at (\i, 0) {};
            %     \node[tensor, font=\footnotesize] (x\i*) at (\i, -0.5) {};
            %     \draw[-] (x\i) -- (x\i*); 
            % \fi
    \foreach  \i in {1,...,6} {
        \pgfmathtruncatemacro{\next}{\i + 1}
            \draw[-] (x\i) -- (x\next);
    }
\end{tikzpicture}.}
\end{align*}
\end{itemize}

\section*{Effect of bond dimension on classification performance} \label{app:bonddim}

To analyze the effect of the bond dimension \( D \) on classification accuracy, we conducted an experiment in which we varied \( D \) while keeping all other hyperparameters fixed.

% \begin{table}[h]
%     \centering
%     \begin{tabular}{c|c}
%         \( D \) & Classification Accuracy \\
%         \hline
%         1  & 0.9455 \\
%         2  & 0.9906 \\
%         3  & 0.9905 \\
%         4  & 0.9895 \\
%         10 & 0.9942 \\
%         15 & 0.9930 \\
%         30 & 0.9917 \\
%         60 & 0.9943 \\
%         100 & 0.9933 \\
%     \end{tabular}
%     \caption{Classification accuracy as a function of bond dimension \( D \), for Fourier embedding on the 2D spiral dataset.}
%     \label{tab:bonddim}
% \end{table}

\begin{figure}[!h]
    \centering
    \includegraphics[width=0.7\linewidth]{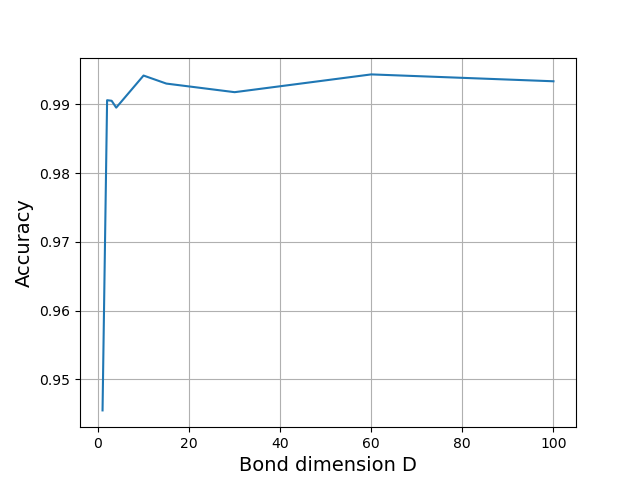}
    \caption{Classification accuracy as a function of bond dimension \( D \), for Fourier embedding on the 2D spiral dataset.}
    \label{fig:bonddim_plot}
\end{figure}

The results presented in Fig.~\ref{fig:bonddim_plot} indicate that classification accuracy initially improves with \( D \) but saturates beyond \( D \approx 4 \). This suggests that for this dataset, the expressive power provided by higher bond dimensions does not yield further improvements. 

A possible explanation is that this dataset has low intrinsic dimensionality and lacks long-range correlations, which are typically modeled by increasing \( D \). In problems with strong correlations between distant variables, a larger bond dimension would be necessary to capture these dependencies. However, in this case, the dataset structure does not demand high expressivity, making smaller bond dimensions sufficient.

\section*{Impact of binning on accuracy and computational cost} \label{app:binning}

The number of bins used to discretize the input space directly impacts both approximation accuracy and computational efficiency. A finer discretization, i.e., increasing the number of bins, leads to a more precise numerical approximation, reducing the squared error. However, it also increases computational cost due to the higher resolution required for numerical integration and sampling.

To analyze the impact of binning on both approximation accuracy and computational efficiency, we conducted an experiment where we sampled from the identity matrix $I_d$ with a fixed physical dimension $d = 10$ using the Fourier embedding. The quantile value was set to $\nu = 0.5$, which allowed us to know the theoretical expected result for the sample $x_1 = 0.5$ and compute the squared loss against the true value.

\begin{table}[!h]
    \centering
    \begin{tabular}{c|c|c}
        \textbf{Number of bins} & \textbf{Squared error} & \textbf{Computation time (s)} \\
        \hline
        \(10^1\)  & \(3.09 \times 10^{-3}\)  & \(4.68 \times 10^{-4}\) \\
        \(10^2\)  & \(2.55 \times 10^{-5}\)  & \(2.26 \times 10^{-4}\) \\
        \(10^3\)  & \(2.51 \times 10^{-7}\)  & \(2.29 \times 10^{-4}\) \\
        \(10^4\)  & \(2.50 \times 10^{-9}\)  & \(1.17 \times 10^{-3}\) \\
        \(10^5\)  & \(2.51 \times 10^{-11}\) & \(4.86 \times 10^{-3}\) \\
        \(10^6\)  & \(2.57 \times 10^{-13}\) & \(5.87 \times 10^{-2}\) \\
        \(10^7\)  & \(3.55 \times 10^{-15}\) & \(6.75 \times 10^{-1}\) \\
    \end{tabular}
    \caption{Effect of the number of bins on squared error and computation time.}
    \label{tab:binning}
\end{table}

\begin{figure}[!h]
    \centering
    \includegraphics[width=0.7\linewidth]{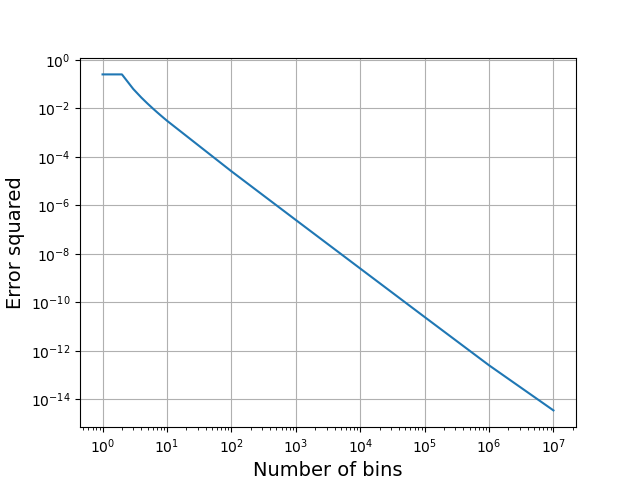}
    \caption{Squared error as a function of the number of bins.}
    \label{fig:binning_error} 
\end{figure}
% BZ: Perhaps it would make sense to have the Y axis in log scale and increase the font in the figure. Similarly in the next figure.

\begin{figure}[h!]
    \centering
    \includegraphics[width=0.7\linewidth]{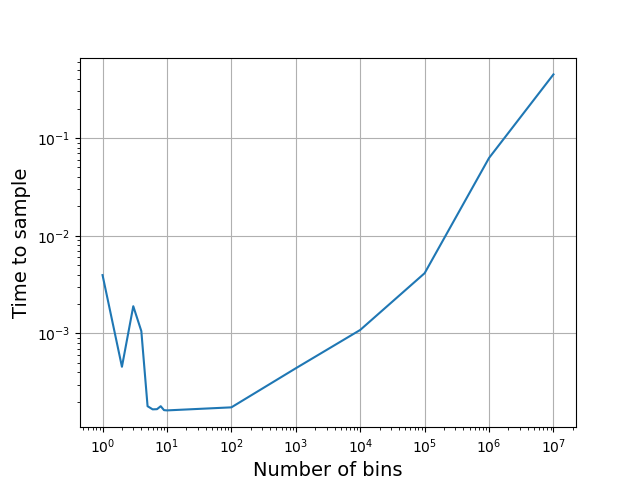}
    \caption{Computation time for a sample as a function of the number of bins.}
    \label{fig:binning_time}
\end{figure}

Figure~\ref{fig:binning_error} illustrates how the squared error decreases exponentially with the number of bins, demonstrating that beyond $10^3$ bins, further improvements become negligible. Conversely, Figure~\ref{fig:binning_time} shows that computational time initially remains low but begins increasing significantly beyond $10^5$ bins. This reflects the increased computational burden required to process finer binning resolutions.

Thus, after analyzing the complete data, presented in Table~\ref{tab:binning}, we selected $1000$ bins as the optimal balance, where the squared error is sufficiently low and computational time remains efficient. This choice ensures numerical stability while avoiding unnecessary overhead.

% \ifCLASSOPTIONcaptionsoff
%   \newpage
% \fi

\section*{Dataset Generation} \label{app:datasets}
In this appendix, we provide details on the datasets used in our experiments.

\subsection{2D Spiral Dataset}
The spiral dataset was generated by sampling \( N = 8000 \) points along two spirals, with angles drawn from a uniform distribution and perturbed with Gaussian noise. The data was then normalized to fit within \([0,1]^2\).

The dataset was generated using the following Python script:
\begin{lstlisting}
import numpy as np  

N = 8000  
theta = np.sqrt(np.random.rand(N)) * 2 * np.pi  

# First spiral
r_a = 2 * theta + np.pi  
x_a = np.array([np.cos(theta) * r_a,
                np.sin(theta) * r_a]).T  
x_a += np.random.randn(N, 2)  

# Second spiral
r_b = -2 * theta - np.pi  
x_b = np.array([np.cos(theta) * r_b,
                np.sin(theta) * r_b]).T  
x_b += np.random.randn(N, 2)  

# Normalize and stack
x_a, x_b = x_a / 20., x_b / 20.  
x = np.vstack((x_a, x_b))  
x = (x - x.min()) / (x.max() - x.min())  
y = np.vstack((np.zeros((N,1)),
               np.ones((N,1))))  
\end{lstlisting}

\subsection{Two Moons Dataset}
The Two Moons dataset was generated using the \texttt{sklearn.datasets.make\_moons} function with added noise.

\begin{lstlisting}
from sklearn.datasets import make_moons  

X, y = make_moons(n_samples=2000, noise=0.1)  
\end{lstlisting}

\subsection{Iris Dataset}
The Iris dataset was obtained using \texttt{sklearn.datasets.load\_iris}. Features were normalized before training.

\begin{lstlisting}
from sklearn.datasets import load_iris  

data = load_iris()  
X, y = data.data, data.target  

# Normalize features
X = (X - X.min()) / (X.max() - X.min())  
\end{lstlisting}

\end{document}